\documentclass{jpp}
\usepackage{amsmath,amssymb}
\usepackage{graphicx,color}
\usepackage[latin1]{inputenc}
\usepackage{natbib}
\usepackage{float}
\usepackage{tabularx}
    \newcolumntype{L}{>{\raggedright\arraybackslash}X}

\usepackage{verbatim}

\newcommand{\bB}{\mathbf{B}}

\newcommand{\bv}{{\mathbf{v}}}

\newcommand{\dd}{{\rm d}}
\newcommand{\rcoor}{{r}}

\def\iotabar{\lower3pt\hbox{$\mathchar'26$}\mkern-7mu\iota}
\newcommand {\aplt} {\ {\raise-.5ex\hbox{$\buildrel<\over\sim$}}\ }

\newcommand{\eq}[1]{(\ref{#1})}
\newcommand{\bun}{\hat{\mathbf{b}}}

\newcommand{\dotcross}{ \raise 0.65ex\hbox{${\scriptstyle {{_{\displaystyle \cdot}}\atop\times}}$} }
\newcommand{\crossdot}{ \raise 0.5ex\hbox{${\scriptstyle {{_\times}\atop{\displaystyle \cdot}}}$} }

\newsavebox{\astrutbox}
\sbox{\astrutbox}{\rule[-5pt]{0pt}{20pt}}

\mathchardef\varLambda="0103

\title[Electrostatic potential on stellarator magnetic surfaces]{Electrostatic potential variations on stellarator magnetic surfaces in low collisionality regimes}

\author[I. Calvo, J.~L. Velasco, F.~I. Parra, J.~ A. Alonso, J.~M. Garc\'ia-Rega\~{n}a]{
Iv\'an Calvo$^1$, Jos\'e Luis Velasco$^1$, F\'elix I. Parra$^{2,3}$, J. Arturo Alonso$^1$ and Jos\'e Manuel Garc\'ia-Rega\~{n}a$^1$
}
\affiliation{$^1$Laboratorio Nacional de Fusi\'on, CIEMAT, 28040 Madrid, Spain\\[2pt]
$^2$Rudolf Peierls Centre for Theoretical Physics, University of Oxford, Oxford, OX1 3PU, 
UK\\[2pt]
$^3$Culham Centre for Fusion Energy, Abingdon, OX14 3DB, UK}

\begin{document}

\maketitle

\begin{abstract}
The component of the neoclassical electrostatic potential that is non-constant on the magnetic surface, that we denote by $\tilde\varphi$, can affect radial transport of highly charged impurities, and this has motivated its inclusion in some modern neoclassical codes. The number of neoclassical simulations in which $\tilde\varphi$ is calculated is still scarce, partly because they are usually demanding in terms of computational resources, especially at low collisionality. In this paper the size, the scaling with collisionality and with aspect ratio, and the structure of $\tilde\varphi$ on the magnetic surface are analytically derived in the $1/\nu$, $\sqrt{\nu}$ and superbanana-plateau regimes of stellarators close to omnigeneity; i. e. stellarators that have been optimized for neoclassical transport. It is found that the largest $\tilde\varphi$ that the neoclassical equations admit scales linearly with the inverse aspect ratio and with the size of the deviation from omnigeneity. Using a model for a perturbed omnigeneous configuration, the analytical results are verified and illustrated with calculations by the code \texttt{KNOSOS}. The techniques, results and numerical tools employed in this paper can be applied to neoclassical transport problems in tokamaks with broken axisymmetry.
\end{abstract}

\section{Introduction}
\label{sec:introduction}

The transport of impurities in three-dimensional toroidal magnetic fields has attracted much attention from the stellarator community. In stellarators, the accumulation of impurities in the center of the confinement region  often limits the discharge duration and is considered to be a potential problem for the development of future reactors. In the framework of neoclassical theory, this accumulation has generally been explained with the inward convection caused by the typically negative radial electric field acting on the highly charged ions in the absence of a so-called ``temperature screening'' effect in non-axisymmetric systems\footnote{Recently, both the prevalence of the radial electric field in the transport of impurities \ and the absence of impurity screening in three-dimensional magnetic fields have been brought into question for several collisionality regimes \citep{VelascoNF2017, HelanderPRL2017}.
}. In a qualitative sense, these expectations are consistent with the general trend observed in the impurity confinement time \citep{BurhennNF2009}, although there are remarkable experimental observations of outward impurity transport (see, for example, \citet{McCormickPRL2002, IdaPoP2009}). This outward impurity flux is still poorly understood. On a quantitative level, it is difficult to determine whether the observed impurity fluxes agree with calculations relying on the solution of approximate versions of the drift-kinetic equation, for these comparisons are often fragmented, dealing with a reduced number of plasma profiles and based on different measuring techniques. The fundamental output of neoclassical modeling of impurity transport is 
the spatially resolved radial particle flux of a charge state of a certain impurity species,  and it is often the case that this very quantity is not experimentally accessible, which complicates the direct quantitative comparisons.
  
The kinetic modeling of impurity dynamics and transport has undergone recent improvements with the inclusion of terms that had previously been considered of secondary importance \citep{GarciaPPCF2013}. These terms model the modification of the trajectories of impurity ions by the component of the electrostatic potential that is non-constant on the flux surface, that we denote by $\tilde\varphi$ and that defines the component of the electric field that is tangent to the flux surface. Specifically, this tangential electric field produces a radial $E\times B$ drift that advects impurities on the flux surface. The effect of $\tilde\varphi$ on impurity transport is larger for impurities with higher electric charge.

A previous step to the assessment of the impact of $\tilde\varphi$ on radial impurity transport is the neoclassical calculation of $\tilde\varphi$ itself\footnote{As a matter of fact, this has brought up the relevance of the measurement of $\tilde\varphi$~\citep{PedrosaNF2015} and its impact on impurity density inhomogeneities \citep{Arevalo2014, Alonso2016}.}, which has typically been ignored in standard neoclassical simulations. The modern codes \texttt{EUTERPE} \citep{Kornilov2005} and \texttt{SFINCS} \citep{Landreman2014} can compute it. In \cite{GarciaNF2017}, results of $\tilde\varphi$ from both codes for  different stellarators and plasma regimes are presented. These extended neoclassical models have started to be used to revisit some of the previous calculations of neoclassical impurity fluxes and have shown substantial deviations (with respect to codes that do not include the effect of $\tilde\varphi$) for some machines and plasma conditions \citep{GarciaNF2017, Mollen2018}. Recently, the strong effect of $\tilde\varphi$ on the neoclassical transport of highly charged collisional impurities in stellarators has been analytically proven in \citep{Calvo2018}.

Given the relevance of correctly determining $\tilde\varphi$, it is important to understand the equations that have to be solved to calculate it, and to derive and discuss analytically some features of the solutions to these equations. In this paper we do so for main ions in the $1/\nu$, $\sqrt{\nu}$ and superbanana-plateau low collisionality regimes (see \citet{Mynick1984} for estimations of the effect of $\tilde\varphi$ on main ion transport in low collisionality regimes under some simplified assumptions). Low collisionality regimes are the pertinent ones for the main ions in the core of reactor-relevant stellarator plasmas and, in addition, in these regimes $\tilde\varphi$ is large and has significant impact on impurity transport. We will carry out such analytical discussion by using the techniques and results of \citet{CalvoPPCF2017}, where we studied the radial neoclassical transport of low collisionality ions in stellarators close to omnigeneity. Omnigeneous magnetic fields~\citep{Hall1975, Cary1997b, Parra2015} are those for which the orbit-averaged radial magnetic drift vanishes for all trapped particles. Omnigenous stellarators can be said to be perfectly optimized from a neoclassical point of view, exhibiting transport levels similar to those of a tokamak. In \citet{CalvoPPCF2017}, expansions in large aspect ratio were not considered. In this article, we expand in small inverse aspect ratio (in the end, all stellarators in operation today have large aspect ratio \citep{Beidler2011}). We will see that the large aspect ratio expansion introduces a number of subtleties.

Let us emphasize that closeness to omnigeneity is not a mere academic requirement. It is well-known \citep{Galeev1979, Ho1987} that, for sufficiently low ion collisionality, tangential drifts have to be kept in the stellarator drift-kinetic equation. If the radial electric field is not small and the aspect ratio is large, one can show that the drift-kinetic equation is radially local and that the tangential magnetic drift is negligible compared to the tangential $E\times B$ drift. But if one of these assumptions fails (that is, if the radial electric field is small or the stellarator is compact), the drift-kinetic equation becomes radially global (with obvious negative consequences for confinement, because then the radial excursion associated with a particle orbit can be as large as the stellarator minor radius) at small collisionality values unless the stellarator is close to omnigeneity. If the magnetic configuration is close to being omnigeneous and the deviation from omnigeneity has small spatial gradients, the set of equations consisting of the drift-kinetic equation and the quasineutrality equation can be rigorously proven to be linear and radially local \citep{CalvoPPCF2017}. Not only do the resulting equations include the effect of $\tilde\varphi$, but also the effect of the tangential magnetic drift. As we have pointed out, the tangential magnetic drift has to be taken into account to describe the transport of the main ions at very low collisionality in compact stellarators and in situations of small radial electric field even if the aspect ratio is large. The importance of the tangential magnetic drift in the latter case is a subject of active research \citep{Matsuoka2015}. In this paper, we will pay special attention to the behavior of $\tilde\varphi$ in regimes where the tangential magnetic drift has to be retained.

Apart from discussing $\tilde\varphi$ analytically in the $1/\nu$, $\sqrt{\nu}$ or superbanana-plateau regimes, we will verify and illustrate the results by comparing them with \texttt{KNOSOS} \citep{Velasco2018}, a newly developed bounce-averaged code that solves the drift-kinetic and quasineutrality equations derived in \citet{CalvoPPCF2017} (for the time being, \texttt{KNOSOS} uses a pitch-angle scattering collision operator), and therefore retains the tangential magnetic drift (among the two codes mentioned some paragraphs above, \texttt{EUTERPE} does not retain the tangential magnetic drift, whereas a way to try to retain it in \texttt{SFINCS} has been recently reported in \citet{Paul2017}). Finally, a remark on the scope of this paper is in order. Whereas the expansion around omnigeneity should be considered as a realistic approach to the analysis of magnetic configurations that aspire to confine sufficiently well to be the basis of stellarator reactor designs, our present treatment is limited by the requirement of small gradients of the non-omnigeneous perturbation in \citet{CalvoPPCF2017}. It would be important to address the relaxation of this condition in future research. Some ideas about the kind of complications introduced by large gradients can be found in \citet{Calvo14, Calvo15} in the setting of perturbed quasisymmetric configurations. Stellarators close to quasisymmetry where the non-quasisymmetric perturbation has small gradients were treated in detail in \cite{Calvo13}. Finally, we note that a tokamak with (slightly) broken axisymmetry can be viewed as a particular case of a magnetic configuration close to omnigeneity, and therefore the tools and results of this paper can be directly applied to problems such as the calculation of tokamak neoclassical toroidal viscosity.
 
The rest of the paper is organized as follows. In Section \ref{sec:basicequations}, after defining the coordinates that will be employed in the theoretical derivations (subsection \ref{sec:phase_space_coordinates}) and introducing the notion of omnigeneity (subsection \ref{sec:omnigeneity}), we present in subsection \ref{sec:driftkinetic_and_quasineutrality_eqs} the neoclassical equations that determine $\tilde\varphi$ in low collisionality regimes of stellarators close to omnigeneity when the deviation from omnigeneity has small spatial gradients. We will also see how $\tilde\varphi$ scales with the size of the deviation from omnigeneity.  In subsection \ref{sec:large_aspect_ratio}, and for large aspect ratio, the scaling with aspect ratio of some quantities entering the neoclassical equations is given; these expressions will be useful in later sections. In subsection \ref{sec:model_for_B}, the magnetic field close to omnigeneity that we will use for the numerical examples is described. With this magnetic field, and taking reactor-relevant parameter values, in Section \ref{sec:realistic_numerical_example_varphi1} we give a plot (see figure \ref{fig:varphi1_vs_nustar_realistic_values}) calculated by \texttt{KNOSOS} showing the dependence of the size of the non-constant component of the electrostatic potential on the collisionality and the radial electric field. The objective of the next section, Section \ref{sec:calculation_varphi1}, is to explain mathematically the different regimes observed in plots like that in figure \ref{fig:varphi1_vs_nustar_realistic_values}. Hence, in Section \ref{sec:calculation_varphi1} we discuss analytically the features of $\tilde\varphi$ in the $1/\nu$, $\sqrt{\nu}$ and superbanana-plateau asymptotic regimes as implied by the equations given in subsection \ref{sec:driftkinetic_and_quasineutrality_eqs} (as explained in \citet{CalvoPPCF2017}, those equations are not valid for lower collisionality regimes like the superbanana or $\nu$ regimes, the essential reason being that for such low collisionality values one has to take into account the radial displacement of the trajectories). In particular, we will comment on the scaling of $\tilde\varphi$ with collisionality and aspect ratio, as well as on its behavior under stellarator symmetry transformations. Detailed numerical checks of the analytical results with \texttt{KNOSOS} are provided throughout this section. Of course, assuming closeness to omnigeneity is not required if one is only interested in studying the $1/\nu$ and $\sqrt{\nu}$ regimes (the former exists in any stellarator and the latter exists in large aspect ratio stellarators with non-small radial electric field), but the assumption is necessary to be able to study, in the same framework, those two collisionality regimes together with the superbanana-plateau regime. In Section \ref{sec:conclusions}, we give the conclusions. In particular, table \ref{tab:table} condenses the main analytical results derived in the paper. Finally, we point out that, for simplicity, throughout the paper we assume that the electrons are adiabatic. This is a good approximation in many cases, but its accuracy certainly depends on the particular collisionality regimes in which the main ions and the electrons are, and situations in which the electrons have to be treated kinetically are not uncommon (see, e.g. \citet{VelascoNF2017, Garcia2018}). In these situations, a kinetic treatment of the electrons is also needed to calculate $\tilde\varphi$.

\section{Low collisionality neoclassical equations in stellarators close to omnigeneity}
\label{sec:basicequations}

\subsection{Phase space coordinates}
\label{sec:phase_space_coordinates}

In the kinetic description of low collisionality regimes, it is useful to employ spatial coordinates $\{\rcoor,\alpha, l\}$, defined as follows. The radial coordinate $\rcoor \in[0,a]$ labels the magnetic surface and has dimensions of length, $a$ being the minor radius of the stellarator. For given $\rcoor$, the angular coordinate $\alpha\in[0,2\pi)$ selects a magnetic field line on the surface. The coordinate $l \in [0,l_{\rm max}(\rcoor,\alpha))$ is the arc length along the field line. In these coordinates the magnetic field reads
\begin{equation}\label{eq:B}
\bB = \Psi(\rcoor)\nabla\rcoor\times\nabla\alpha.
\end{equation}
As for the meaning of the flux function $\Psi(\rcoor)$, it is easy to check that  $2\pi \int_0^{\rcoor_0} \Psi(\rcoor)\dd\rcoor$ equals the magnetic flux across a surface defined by $l={\rm const.}$ and $0\le \rcoor \le \rcoor_0$.

In these coordinates the flux-surface average of a function $f(\rcoor,\alpha,l)$ is defined as
\begin{equation}\label{eq:def_fluxsurfaceaverage}
\langle f \rangle (\rcoor) = V'(\rcoor)^{-1}\int_{0}^{2\pi} \dd\alpha \int_0^{l_{\rm max}}\dd l \, \Psi B^{-1} f,
\end{equation}
where $\Psi B^{-1} = [(\nabla\rcoor\times\nabla\alpha)\cdot\nabla l]^{-1}$ is the volume element (the coordinates $\{\rcoor,\alpha,l\}$ are chosen such that $\Psi > 0$), $B(\rcoor,\alpha,l)$ is the magnitude of $\bB$ and
\begin{equation}\label{eq:def_volume}
V'(\rcoor) = \int_0^{2\pi} \dd\alpha \int_0^{l_{\rm max}}\dd l \, \Psi B^{-1},
\end{equation}
with $V(\rcoor)$ being the volume enclosed by the surface $\rcoor$. We will use primes to denote differentiation with respect to $\rcoor$.

In velocity space, we employ coordinates $\{v,\lambda,\sigma\}$, where $v$ is the magnitude of the velocity $\bv$, $\lambda = v_\perp^2 / (B v^2)$ is the pitch-angle coordinate, $v_\perp$ is the component of $\bv$ perpendicular to $\bB$ and $\sigma = v_{||} / |v_{||}|$ is the sign of the parallel velocity, with
\begin{equation}
v_{||}(\rcoor,\alpha,l,v,\lambda,\sigma) = \sigma v \sqrt{1- \lambda B(\rcoor,\alpha,l)}\, .
\end{equation}

Integrals over velocity space of gyrophase-independent functions (all functions in this paper are independent of gyrophase) read, in these coordinates,
\begin{equation}
\int (\cdot)\dd^3v \equiv \sum_\sigma\int_0^\infty\dd v\int_0^{B^{-1}} \dd\lambda \, \frac{\pi B v^3}{|v_{||}|} (\cdot) .
\end{equation}

\subsection{Perturbed omnigeneous magnetic fields}
\label{sec:omnigeneity}

\begin{figure}
\centering
\includegraphics[width=0.6\textwidth]{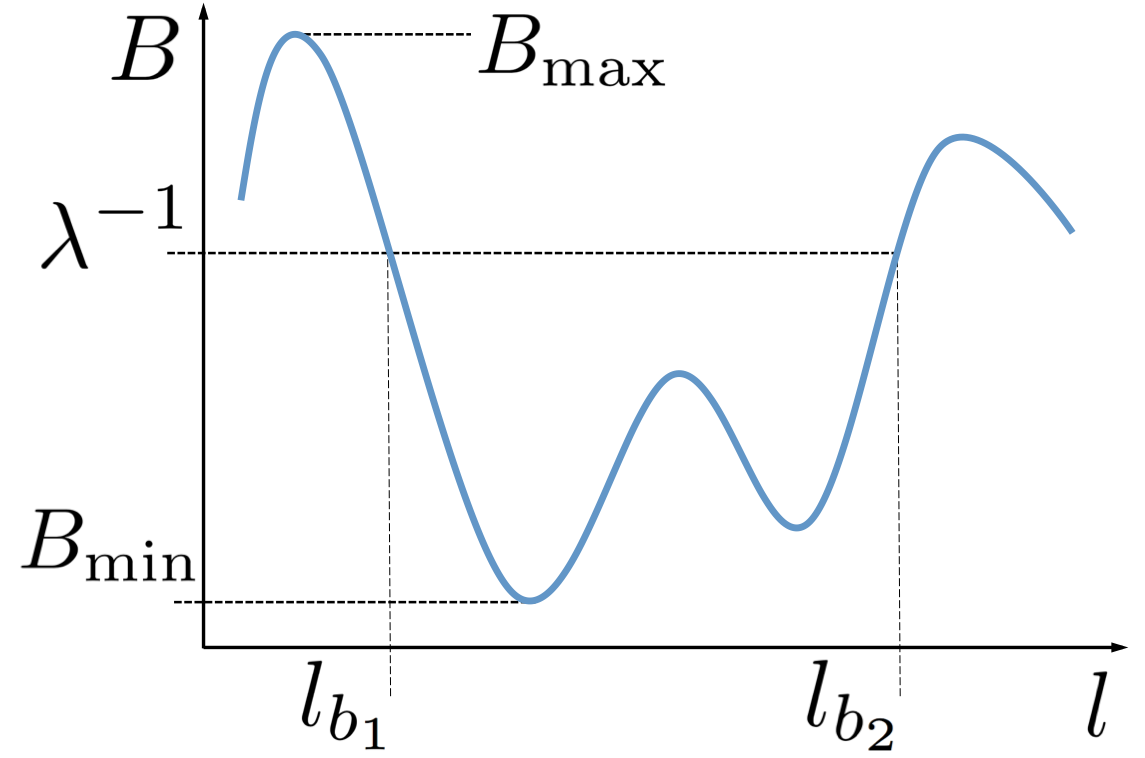}
\caption{Cartoon showing some of the quantities (defined in the text) employed to describe a trapped trajectory.}
\label{fig:trapped_particle}
\end{figure}

Trapped trajectories are those with $\lambda \in[B_{\rm max}^{-1}, B_{\rm min}^{-1}]$, where $B_{\rm max}(\rcoor)$ and $B_{\rm min}(\rcoor)$ are, respectively, the maximum and minimum values of $B$ on the flux surface. For each trapped trajectory,  the second adiabatic
invariant reads
\begin{equation}
J(\rcoor,\alpha,v,\lambda) = 2\int_{l_{b_1}}^{l_{b_2}} |v_{||}| \dd l.
\end{equation}
Here, $l_{b_1}$ and $l_{b_2}$ are the bounce 
points of the trapped orbit, obtained by solving in $l$ the equation $1-\lambda B(\rcoor,\alpha,l) = 0$ (see figure \ref{fig:trapped_particle} for a cartoon showing some of the quantities that describe trapped trajectories).

The importance of $J$ comes from the relations
\begin{eqnarray}
 && \frac{2}{\tau_b}
 \int_{l_{b_1}}^{l_{b_2}}\frac{1}{|v_{||}|}\bv_M
\cdot\nabla\rcoor \, \dd l
=
\frac{m}{Z e\Psi \tau_b}\partial_\alpha J, \label{eq:integral_radial_vM}
\\
&& \frac{2}{\tau_b}
 \int_{l_{b_1}}^{l_{b_2}}\frac{1}{|v_{||}|}\bv_M
\cdot\nabla\alpha \, \dd l
=
-\frac{m}{Z e\Psi \tau_b}\partial_\rcoor J, \label{eq:integral_tangential_vM}
\end{eqnarray}
where
\begin{equation}
\bv_M = \frac{m v^2(1-\lambda B)}{Ze B}\bun\times(\bun\cdot\nabla\bun) + \frac{m v^2\lambda}{2ZeB}\bun\times\nabla B
\end{equation}
is the magnetic drift, $\bun = B^{-1} \bB$, $m$ is the particle mass, $Ze$ is the particle charge (we assume $Z\sim 1$), $e$ is the proton charge and 
\begin{equation}
\tau_b(\rcoor,\alpha, v,\lambda) = 2
\int_{l_{b_{1}}}^{l_{b_{2}}} |v_{||}|^{-1} \dd l
\end{equation}
is the bounce time; i.e. the time that the trapped particle needs to complete its orbit. Hence, $\partial_\alpha J$ and $\partial_\rcoor J$ contain the information on the motion of the trapped particle in the directions perpendicular to $\bB$, averaged over the motion in the direction parallel to $\bB$.

The magnetic field is called omnigeneous~\citep{Cary1997b, Parra2015} if the average radial magnetic drift vanishes for all trapped particles. Due to \eq{eq:integral_radial_vM}, this is equivalent to saying that the second adiabatic invariant is independent of $\alpha$, $\partial_\alpha J = 0$, for all trapped trajectories.

From the definition of omnigeneity one can deduce that, for any function $q$ that depends on $\alpha$ and $l$ only through an omnigeneous magnetic field $B$, one has~\citep{Cary1997b, Helander2009, CalvoPPCF2017}
\begin{equation}\label{eq:definitionomnigEquiv}
\partial_\alpha \int_{l_{b_1}}^{l_{b_2}}q(\rcoor,v,\lambda,B(\rcoor,\alpha,l))
\dd l = 0.
\end{equation}
This useful property is employed in the derivation of the neoclassical equations of subsection \ref{sec:driftkinetic_and_quasineutrality_eqs} and it will also be used in the analysis of their solutions.

In this paper, we will carry out the calculations in magnetic configurations that can be written as an exactly omnigeneous magnetic field $\bB_0$ plus a perturbation $\delta \bB_1$,
\begin{equation}
\bB = \bB_0 + \delta \bB_1,
\end{equation}
where $B_0 := | \bB_0 | \sim | \bB_1 | =: B_1$ and $0 \le \delta \ll 1$. As advanced in Section \ref{sec:introduction}, and following \citet{CalvoPPCF2017}, we restrict ourselves to cases in which the characteristic scale of variation of $B_1$ on the flux surface is not much smaller than that of $B_0 $. Specifically, we require $|\partial_\alpha B_1|/|\partial_\alpha B_0| \sim |\partial_l B_1|/|\partial_l B_0|  \ll \delta^{-1}$. Thus, in terms of an expansion in $\delta\ll 1$, we can assume that $B_0$ and $B_1$ have a similar characteristic variation length on the flux surface, that we can take to be $R_0$, the major radius of the stellarator. This is what we mean when we say that the deviation from omnigeneity has small gradients. In particular, deviations with small gradients can deform the omnigeneous magnetic wells, but not create new wells.

\subsection{Drift-kinetic and quasineutrality equations in stellarators close to omnigeneity}
\label{sec:driftkinetic_and_quasineutrality_eqs}

In drift kinetics~\citep{Hazeltine73, Parra11, Calvo13}, the characteristic frequency of particle motion in the direction parallel to $\bB$ is $O(v_t / R_0)$, whereas the motion perpendicular to the magnetic field has a much smaller typical frequency $O(\rho_* v_t / R_0)$, where $\rho_* = \rho / R_0 \ll 1$ is the normalized Larmor radius, with $\rho = v_t / \Omega$ and $\Omega = Ze B/ m$. Low collisionality regimes\footnote{From here on, we understand that quantities without a subindex specifying the species refer to ions. Electron quantities will always include a subindex $e$. In some ion quantities where some ambiguity might exist, as in $\nu_{ii}$, we include subindices.} are defined by $\nu_{*} \ll 1$, where $\nu_* = \nu_{ii} R_0 / v_t$ is the ion collisionality, $\nu_{ii}$ is the ion-ion collision frequency and $v_t = \sqrt{T/m}$ is the ion thermal speed. The condition $\nu_{*} \ll 1$  means that the characteristic parallel streaming frequency is much greater than the collision frequency $\nu_{ii}$, and therefore the drift-kinetic ion equation can be averaged over the parallel motion. In this subsection, we give the neoclassical equations for low collisionality plasmas in stellarators close to omnigeneity where the deviation from omnigeneity has small gradients, but we do not include their derivation. For that, we refer the reader to \citet{CalvoPPCF2017}. No assumptions are made about the aspect ratio of the stellarator yet.

For $\delta \ll 1$, one finds that the electrostatic potential, $\varphi$, is a flux function to lowest order,
\begin{equation}
\varphi(\rcoor,\alpha,l) = \varphi_0(\rcoor) + \delta\varphi^{(1)}(\rcoor,\alpha,l) + \dots
\end{equation}
The component of the electrostatic potential that is non-constant on the flux surface, denoted in Section \ref{sec:introduction} by $\tilde\varphi$, is therefore given by
\begin{equation}
\tilde\varphi = \varphi - \varphi_0 = \delta\varphi^{(1)} + \dots
\end{equation}

In an expansion in $\delta$, the ion distribution function, $F$, can be written as
\begin{equation}
F = F_M - \frac{Z e \delta \varphi^{(1)}}{T}F_M + \delta g^{(1)} + \dots,
\end{equation}
where $F_M$ is a Maxwellian distribution constant on flux surfaces,
\begin{equation}
F_M(\rcoor,v) = 
n(\rcoor)\left(\frac{m}{2\pi T(\rcoor)}\right)^{3/2}
\exp\left(-\frac{m v^2}{2T(\rcoor)}\right),
\end{equation}
and $n$ and $T$ are the ion density and temperature. The non-adiabatic perturbation to the Maxwellian distribution, $g^{(1)}(\rcoor,\alpha,l,v,\lambda)$, is independent of $l$ and $\sigma$, and vanishes for passing trajectories of the omnigenous field, $0 \le \lambda < B_{0,{\rm max}}^{-1}$, where $B_{0,{\rm max}}(\rcoor)$ is the maximum value of $B_0$ on the flux surface. The function $g^{(1)}$ can be chosen such that
\begin{equation}\label{eq:condition_g1}
\int_0^{2\pi} g^{(1)}
\dd\alpha= 0.
\end{equation}

The equations that determine $g^{(1)}$ and $\varphi^{(1)}$ are radially local and linear. First, one has the bounce-averaged drift-kinetic equation for trapped trajectories,
\begin{eqnarray}\label{eq:DKEtrapped}
-\partial_\rcoor J^{(0)}
\partial_\alpha  g^{(1)}
  + \partial_\alpha J^{(1)} \Upsilon F_{M}
 =
  \sum_\sigma
\frac{Z e\Psi}{m}\,
\int_{l_{b_{10}}}^{l_{b_{20}}}\frac{1}{|v_{||}^{(0)}|}
{C_{ii}^{\ell (0)}[g^{(1)}]} \dd l,
\end{eqnarray}
where\footnote{The presence of $\varphi$ in the expressions for $\partial_\rcoor J^{(0)}$ and $J^{(1)}$ might seem surprising in the light of the discussion on the second adiabatic invariant in subsection \ref{sec:omnigeneity}. When $\varphi$ is non-zero, the average tangential and radial drifts (that now include the corresponding components of the $E\times B$ drift) are obtained by taking derivatives with respect to $\rcoor$ and $\alpha$ keeping ${\cal E} = mv^2/2 + Ze\varphi / m$ and $\mu = v_\perp^2/(2B)$ constant. After this, one can perform the $\delta$ expansion and change to  coordinates $v$ and $\lambda$, obtaining \eq{eq:dJ0/dpsi} and \eq{eq:dJ1/dalpha}. This process is explained step by step in \citet{CalvoPPCF2017}.}
\begin{equation}\label{eq:dJ0/dpsi}
  \partial_\rcoor J^{(0)}
  =
  -\int_{l_{b_{10}}}^{l_{b_{20}}}
\frac{\lambda v\partial_\rcoor B_0 + 2Z e/(mv)\varphi'_0}
{\sqrt{1-\lambda B_0}}\dd l,
\end{equation}
\begin{equation}\label{eq:dJ1/dalpha}
   J^{(1)}
  =
  -\int_{l_{b_{10}}}^{l_{b_{20}}}
\frac{
\lambda v B_1 + 2Z e/(m v)\varphi^{(1)}
}
{\sqrt{1-\lambda B_0}}\dd l, 
\end{equation}
\begin{equation}
\Upsilon = \frac{n'}{n} + \frac{T'}{T}
\left(
\frac{m v^2}{2T}-\frac{3}{2}
\right)
+
\frac{Z e\varphi_0'}{T}
\end{equation}
and $C_{ii}^{\ell}$ is the linearized ion-ion collision operator~\citep{Helander2002_book}. The ion-electron collision term has been neglected because we assume $\sqrt{m_e/m}\ll 1$, with $m_e$ the electron mass. The superindex $(0)$ in $C_{ii}^{\ell (0)}$ means that $B$ has been replaced by $B_0$ in the expression for the collision operator in coordinates $v$ and $\lambda$. Analogously,
\begin{equation}
v_{||}^{(0)}(\rcoor,\alpha,l,v,\lambda,\sigma) = \sigma v \sqrt{1- \lambda B_0(\rcoor,\alpha,l)}\, ,
\end{equation}
and $l_{b_{10}}$, $l_{b_{20}}$ are the solutions of the equation $1-\lambda B_0(\rcoor,\alpha,l) = 0$ for $\lambda \ge B_{0,{\rm max}}^{-1}$. Finally, note that $\partial_\rcoor J^{(0)}$ does not depend on $\alpha$ because $B_0$ is omnigeneous. Sometimes, it will be useful to write $J^{(1)} = J_B^{(1)} + J_\varphi^{(1)}$, where
\begin{equation}\label{eq:J1_B}
J_B^{(1)}
  =
  - \lambda v \int_{l_{b_{10}}}^{l_{b_{20}}}
\frac{
 B_1
}
{\sqrt{1-\lambda B_0}}\, \dd l
\end{equation}
and
\begin{equation}\label{eq:J1_varphi}
J_\varphi^{(1)}
  =
  -\frac{2Ze}{mv}
  \int_{l_{b_{10}}}^{l_{b_{20}}}
\frac{
\varphi^{(1)}
}
{\sqrt{1-\lambda B_0}}\, \dd l.
\end{equation}

Second, one has the quasineutrality equation. In the mass ratio expansion $\sqrt{m_e/m}\ll 1$, the non-adiabatic response of the electrons can be dropped, giving
\begin{eqnarray}\label{eq:quasineutrality_equation}
\left(\frac{Z}{T}+\frac{1}{T_e}\right)\varphi^{(1)} =
\frac{2\pi}{en} 
\int_0^\infty\dd v \int_{B^{-1}_{0,{\rm max}}}^{B_0^{-1}}\dd\lambda
\frac{v^3 B_0}{|v_{||}^{(0)}|}
g^{(1)},
\end{eqnarray}
where $T_e(\rcoor)$ is the electron temperature. Note that the choice \eq{eq:condition_g1} implies that
\begin{equation}
\langle \varphi^{(1)} \rangle = 0,
\end{equation}
where the flux-surface average is taken using $B_0$. In addition, integration of \eq{eq:quasineutrality_equation} over trapped orbits, application of property \eq{eq:definitionomnigEquiv} and of condition \eq{eq:condition_g1} implies
\begin{equation}\label{eq:property_J1_varphi}
\int_0^{2\pi}
J_\varphi^{(1)}\dd\alpha
  = 0.
\end{equation}

Equations \eq{eq:DKEtrapped} and \eq{eq:quasineutrality_equation} are correct for values of the collisionality that include the neoclassical transport regimes known as $1/\nu$, $\sqrt{\nu}$ and superbanana-plateau regimes. In \citet{CalvoPPCF2017}, it was explained why the equations stop to be valid at low enough values of the collisionality, so that they should be modified in case one is interested in describing, for example, the $\nu$ or superbanana regimes. In subsection \ref{sec:large_aspect_ratio}, we explain how some of the quantities entering the neoclassical equations \eq{eq:DKEtrapped} and \eq{eq:quasineutrality_equation} behave when a large aspect ratio expansion is taken. In subsection \ref{sec:model_for_B}, the perturbed omnigeneous magnetic field that will be later used in the numerical examples is described.

\subsection{Conventions and some useful scalings for large aspect ratio stellarators}
\label{sec:large_aspect_ratio}

Stellarators in operation today have large aspect ratio, $\epsilon^{-1} = R_0 / a$. In \citet{CalvoPPCF2017}, the scaling with $\epsilon\ll 1$ was not determined, and we would like to address this refinement in the present paper. In this subsection, some conventions that are convenient for $\epsilon\ll 1$ are explained, and the scaling with $\epsilon$ of several quantities entering \eq{eq:DKEtrapped} and \eq{eq:quasineutrality_equation} is provided for later use.

In a given flux surface, we write magnetic fields in large aspect ratio stellarators close to omnigeneity as
\begin{equation}
B = B_0 + \delta B_1,
\end{equation}
where
\begin{equation}\label{eq:decomposition_B0}
B_0 = B_{00} + \tilde B_0,
\end{equation}
$B_{00}$ is constant and $\tilde B_0 \sim \epsilon B_{00}$. As for the non-omnigenous perturbation, we assume $B_1\sim \epsilon B_{00}$. Note that if the aspect ratio is large, the meaningful measure of the size of the non-omnigeneous perturbation is given by $\delta \sim |B - B_0| / |B_0 - B_{00}|$. This is why we have included a factor $\epsilon$ in our convention for the size of $B_1$. The definition
\begin{equation}
\delta \sim \frac{B - B_0}{\epsilon B_0}
\end{equation}
is valid both for large aspect ratio, $\epsilon\ll 1$, and tight aspect ratio, $\epsilon \sim 1$.

The radial derivative of $B_0$ at the surface of interest will also be needed. In our models, we will assume
\begin{equation}
\partial_\rcoor B_0 = \frac{\tilde{B}_0}{\epsilon R_0}.
\end{equation}

We proceed to explain how different quantities appearing in the neoclassical equations scale with $\epsilon$. We denote by $\Delta_{{\rm trapped}}$ the size (measured in the coordinate $\lambda$) of the region of phase space corresponding to particles trapped in $B_0$. Observing that the difference between the maximum and the minimum values of $B_0$ on the flux surface is $O(\epsilon B_{00})$, we infer that
\begin{equation}
\Delta_{{\rm trapped}}\sim \epsilon B_{00}^{-1}.
\end{equation}
Analogous reasons lead to conclude that the size of the parallel velocity for trapped particles is
\begin{equation}
v_{||}^{(0)}\sim \epsilon^{1/2} v_t.
\end{equation}

The scaling of $\partial_\rcoor J^{(0)}$ depends on the size of the radial electric field, obtained by imposing ambipolarity of the neoclassical radial particle fluxes. For typical values $\varphi'_0 \sim \epsilon^{-1} T / (e R_0)$, the term containing the radial electric field dominates and
\begin{equation}
  \partial_\rcoor J^{(0)}
  \sim \epsilon^{-3/2} v_t,
\end{equation}
where we have assumed that the typical length of the omnigeneous wells is $R_0$. If the radial electric field is small, $|\varphi'_0| \lesssim T / (e R_0)$, then\footnote{See, for example, \cite{Klinger2017}, where radial electric field profiles are shown such that $\varphi'_0 = 0$ at some radial position.}
\begin{equation}
  \partial_\rcoor J^{(0)}
  \sim \epsilon^{-1/2} v_t.
\end{equation}

As for $J^{(1)}$, we have
\begin{equation}
J_B^{(1)} \approx
 - B_{00}^{-1} v \int_{l_{b_{10}}}^{l_{b_{20}}}
\frac{
 B_1
}
{\sqrt{1-\lambda B_0}}\, \dd l
 \sim \epsilon^{1/2}v_t R_0
\end{equation}
whereas
\begin{equation}\label{eq:size_J1_varphi_general}
J_\varphi^{(1)}\sim \epsilon^{-1/2}  v_t R_0 \frac{e \varphi^{(1)}}{T}\, .
\end{equation}
Recall that $\varphi^{(1)}$ is one of the unknowns of the neoclassical equations and its size is to be determined.

The function $\Psi$ appearing in \eq{eq:B} scales as
\begin{equation}
\Psi\sim \epsilon B_{00} R_0.
\end{equation}

Regarding the plasma profile gradients, it is expected that, generically,
\begin{equation}
\Upsilon \sim \frac{1}{\epsilon R_0}.
\end{equation} 

In large aspect ratio stellarators, the collision term on the right-hand side of \eq{eq:DKEtrapped} is dominated by the pitch-angle scattering piece of the collision operator,
\begin{eqnarray}\label{eq:pitch-angle-scatt_large_aspect_ratio}
C_{ii}^{\ell (0)}[g^{(1)}] \approx \nu_\lambda
\frac{ v_{||}^{(0)}}{v^2 B_{00}^2}\partial_\lambda\left(
v_{||}^{(0)}\partial_\lambda g^{(1)}
\right),
\end{eqnarray}
where
\begin{equation}
\nu_\lambda(v) = \frac{3\sqrt{\pi}}{2\tau_{ii}}
\, \frac{{\rm erf}\left(v\sqrt{m / (2T)}\right) - \chi\left(v\sqrt{m/ (2T)}\right)}{\left(v\sqrt{m/ (2T)}\right)^3}
\end{equation}
is the ion-ion collision frequency,
\begin{equation}
\tau_{ii} = 6\sqrt{2} \pi^{3/2} \varepsilon_0^2 m^{1/2} T^{3/2} / (Z^4 e^4 n \ln\Lambda),
\end{equation}
$\ln\Lambda$ is the Coulomb logarithm, $\varepsilon_0$ is the vacuum permittivity,
\begin{equation}
{\rm erf}(x) = \frac{2}{\sqrt{\pi}} \int_0^x e^{-y^2}\dd y
\end{equation}
and
\begin{equation}
\chi(x) = \frac{{\rm erf}(x) - (2x/\sqrt{\pi})\exp(-x^2)
}{2 x^2}\, .
\end{equation}
Employing the estimates above for the scaling of $\Delta_{{\rm trapped}}$ and $v_{||}^{(0)}$ with $\epsilon$, we find
\begin{eqnarray}\label{eq:estimation_collision_operator_small_epsilon}
C_{ii}^{\ell (0)}[g^{(1)}] \sim \frac{\nu_*}{\epsilon}\frac{v_t}{R_0} g^{(1)},
\end{eqnarray}
which is the typical size of the collision term across the trapped region. It is important to realize that the condition for low collisionality ions, that reads $\nu_* \ll 1$ for $\epsilon \sim 1$, becomes\footnote{We point out that the range $\epsilon^{3/2} \ll \nu_* \ll 1$ corresponds to the plateau regime.} $\nu_* \ll \epsilon^{3/2}$ for $\epsilon \ll 1$. This result is derived by comparing the factor in front of $g^{(1)}$ on the right-hand side of \eq{eq:estimation_collision_operator_small_epsilon} with the typical frequency of the parallel particle motion, which is $O(\epsilon^{1/2}v_t/R_0)$ when $\epsilon \ll 1$.

Finally, we point out that (see \eq{eq:pitch-angle-scatt_large_aspect_ratio}), when $\epsilon \ll 1$, we can define
\begin{equation}
\nu_* := \frac{\nu_\lambda(v_t) R_0}{v_t}
\end{equation}
as the expression for the collisionality and
\begin{equation}
\rho_*: = \frac{\sqrt{m T}}{Ze B_{00}R_0}
\end{equation}
for the normalized Larmor radius.

\subsection{Model magnetic field close to omnigeneity}
\label{sec:model_for_B}

\begin{figure}
\centering
\includegraphics[width=0.7\textwidth]{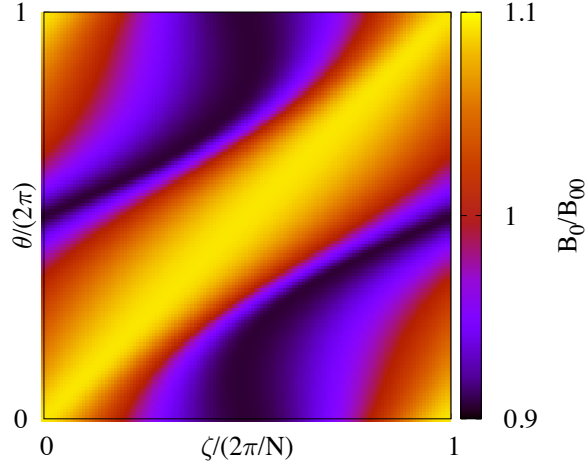}
\caption{Plot of $B_0$ for the model omnigeneous magnetic field described at the beginning of subsection \ref{sec:model_for_B}; that is, $B_0 = B_{00} + \tilde{B}_0$, with $\tilde{B}_0$ of the form \eq{eq:tildeB0_Landreman}. Here, $\epsilon = 0.1$.}
\label{fig:B_0}
\end{figure}

In order to define a specific omnigeneous field $B_0$ for the numerical examples provided in subsequent sections, we follow \citet{Cary1997b} and \citet{Landreman2012}. The description is easier in Boozer coordinates~\citep{Boozer81}. We denote by $\theta$ and $\zeta$ the poloidal and toroidal Boozer angles, respectively. We use the construction explained in section V of  \citet{Landreman2012} with $M=1$ and $N=4$ (which means that the contours of $B_0$ close on themselves after 4 poloidal turns and 1 toroidal turn) and with rotational transform $\iotabar = 1.05$. We take\footnote{In order to avoid any confusion, we point out that the meaning of the symbol $\epsilon$ in our paper and in \citet{Landreman2012} is not exactly the same. What we call $\epsilon$ would be called $\epsilon/(1+\epsilon)$ in \citet{Landreman2012}. What we call $B_{00}(1-\epsilon)$ would be called $\check{B}$ in \citet{Landreman2012}.}
\begin{equation}\label{eq:tildeB0_Landreman}
\tilde{B}_0(\theta,\zeta) = \epsilon B_{00} \cos\eta(\theta,\zeta),
\end{equation}
where the function $\eta$ is obtained as in equation (73) of \citet{Landreman2012}. In figure \ref{fig:B_0} we show a plot of an omnigeneous magnetic field of the form just introduced.

As for the non-omnigeneous perturbation, we take
\begin{equation}\label{eq:defB1}
B_1 = \epsilon B_{00}\cos (2\theta).
\end{equation}

Note that $B = B_0 + \delta B_1$ is symmetric under the transformation $(\theta,\zeta)\mapsto (-\theta,-\zeta)$. Magnetic fields with this property are called \emph{stellarator symmetric} \citep{Dewar1998}. Stellarator designs typically impose this symmetry on the magnetic configuration, and from here on we always assume that our magnetic field satisfies it.

Given Boozer angles $\{\theta,\zeta\}$, there are many different ways to define coordinates $\{\alpha,l\}$. First, one has to choose a closed curve ${\cal C}$ that is not contractible to a point. A simple choice is to take the curve $\zeta = 0$. Then,
\begin{equation}
\alpha := \theta - \iotabar\zeta.
\end{equation}
As for the arc length $l$,
\begin{equation}
l(\alpha,\zeta) = \int_0^{\zeta} \frac{1}{\bun\cdot\nabla\zeta'}\dd\zeta' \,,
\end{equation}
where the integral on the right-hand side is taken at constant $\alpha$ and $l_{\rm max}$ is determined by the intersection of the field line with ${\cal C}$ when it goes all the way around the torus. 

\section{Dependence of $\varphi^{(1)}$ on the collisionality and the radial electric field: a realistic numerical example}
\label{sec:realistic_numerical_example_varphi1}

\begin{figure}
\centering
\includegraphics[width=0.8\textwidth]{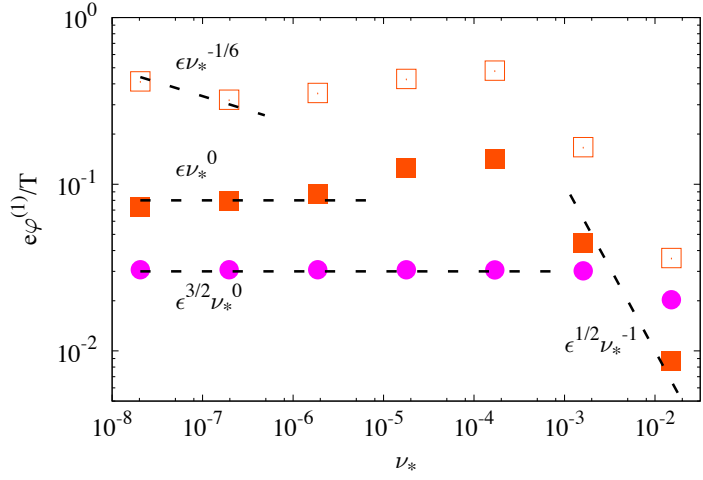}
\caption{Size of $\varphi^{(1)}$ as a function of the collisionality for a plasma with $\rho_* = 1.87\times 10^{-4}$  and magnetic configuration defined at the end of subsection \ref{sec:model_for_B} with $\epsilon = 0.1$. The circles correspond to results with $\varphi'_0 = 2 \epsilon^{-1} T/(eR_0)$. The squares correspond to $\varphi'_0 = 0$. Empty squares give the size of $\varphi^{(1)}$. Full squares give the size of $\varphi^{(1)}$ at  generic points (see the main text for the explanation on the difference between size of $\varphi^{(1)}$ and size of $\varphi^{(1)}$ at generic points). The important difference between the curves consisting of empty and full squares is that, for $\nu_*\ll \rho_*$, the first one behaves as $\nu_*^{-1/6}$ and the second one does not vary with $\nu_*$.}
\label{fig:varphi1_vs_nustar_realistic_values}
\end{figure}

Let us take realistic, stellarator reactor relevant parameters, $\rho_* = 1.87\times 10^{-4}$ and $\epsilon = 0.1$, and let us employ the model magnetic field close to omnigeneity $B = B_0 + \delta B_1$ described in subsection \ref{sec:model_for_B} (see \eq{eq:decomposition_B0}, \eq{eq:tildeB0_Landreman} and \eq{eq:defB1}). As for the profiles, we take $n'/n = - (\epsilon R_0)^{-1}$ and $T' = 0$. We solve \eq{eq:DKEtrapped} and \eq{eq:quasineutrality_equation} with the code \texttt{KNOSOS}, and in figure \ref{fig:varphi1_vs_nustar_realistic_values} we represent the size of $\varphi^{(1)}$ versus $\nu_*$. We plot curves corresponding to $\varphi'_0 = 0$ and to $\varphi'_0 = 2 \epsilon^{-1}T/(e R_0)$. In this plot and in subsequent ones representing $\varphi^{(1)}$ versus $\nu_*$, we typically take the maximum value of $\varphi^{(1)}$ on the flux surface, and this is what we generally understand by ``the size of $\varphi^{(1)}$". However, there are regimes (see subsection \ref{sec:varphi1_sb-p_regime}) in which the scaling of $\varphi^{(1)}$ at special points (lying in a very small spatial region associated with certain resonant trajectories) is different from that of $\varphi^{(1)}$ at generic points, far from the special points. When such regimes exist, we usually plot two curves and distinguish between ``the size of $\varphi^{(1)}$" (see, for example, empty squares in figure \ref{fig:varphi1_vs_nustar_realistic_values}), given by the maximum of $\varphi^{(1)}$ on the flux surface, and ``the size of $\varphi^{(1)}$ at generic points" (see, for example, full squares in figure \ref{fig:varphi1_vs_nustar_realistic_values}). In order to compute the size of $\varphi^{(1)}$ at generic points, we take a spatial average of $\varphi^{(1)}$ in a region where $\varphi^{(1)}$ is positive and which is far from the special points.

 The dependence of $\varphi^{(1)}$ on the collisionality and on the radial electric field shown in figure \ref{fig:varphi1_vs_nustar_realistic_values} is non-trivial. Understanding the different regimes encoded in a plot like figure \ref{fig:varphi1_vs_nustar_realistic_values} is the main subject of this paper. We will identify and study these regimes in Section \ref{sec:calculation_varphi1}. As we will explain below, in order to have very clear numerical scalings in all the asymptotic regimes that \eq{eq:DKEtrapped} and \eq{eq:quasineutrality_equation} admit, in Section \ref{sec:calculation_varphi1} we will employ extremely small values of $\rho_*$ and $\epsilon$ in our numerical checks. The reader that prefers to skip the mathematical derivations of Section \ref{sec:calculation_varphi1} can go directly to table \ref{tab:table} in Section \ref{sec:conclusions} for a collection of the main analytical results.

\section{Asymptotic regimes of $\varphi^{(1)}$ at low collisionality}
\label{sec:calculation_varphi1}

In this section, we discuss the solution $\varphi^{(1)}$ determined by equations \eq{eq:DKEtrapped} and \eq{eq:quasineutrality_equation} in three asymptotic low collisionality regimes. Namely, the $1/\nu$ regime, the $\sqrt{\nu}$ regime and the superbanana-plateau regime (the names of the regimes are due to the scaling of the radial neoclassical fluxes with collisionality, not necessarily to the scaling of $\varphi^{(1)}$, as we will explain below). The subsection devoted to each regime will start with a discussion valid for arbitrary aspect ratio. Then, the results will be particularized for large aspect ratio stellarators and numerical examples of the most relevant analytical results will follow. In the course of the analytical derivations, it will become clear that in order to clearly distinguish all the asymptotic regimes that equations \eq{eq:DKEtrapped} and \eq{eq:quasineutrality_equation} admit, the quantities $\rho_* $, $\rho_*/\epsilon^{1/2}$ and $\rho_*/\epsilon$ need to be sufficiently separated from each other. This is why, in what follows, we will take artificially small values for $\rho_*$ and $\epsilon$, so that there is enough room between those quantities in our numerical examples. In these examples, we will always take $\rho_* = 1.1\times 10^{-12}$, $n'/n = -(\epsilon R_0)^{-1}$ and $T' = 0$, and we will always employ the model magnetic field close to omnigeneity defined in subsection \ref{sec:model_for_B} (recall equations \eq{eq:decomposition_B0}, \eq{eq:tildeB0_Landreman} and \eq{eq:defB1}). We will use different values of $\epsilon$ and $\varphi'_0$ that will be specified where appropriate.

\subsection{The $1/\nu$ regime}
\label{sec:varphi1_1overnu_regime}

For tight aspect ratio stellarators, the $1/\nu$ regime corresponds to $\rho_* \ll \nu_*\ll 1$. In this situation, the first term on the left-hand side of \eq{eq:DKEtrapped} is negligible compared to the term on the right-hand side,
\begin{equation}
\left|
\partial_\rcoor J^{(0)} \partial_\alpha  g^{(1)}
\right|
\ll
\left|
\sum_\sigma
\frac{Z e\Psi}{m}\,
\int_{l_{b_{10}}}^{l_{b_{20}}}\frac{\dd l}{|v_{||}^{(0)}|}
{C_{ii}^{\ell (0)}\left[g^{(1)}\right]}
\right|
.
\end{equation}

The correction to the Maxwellian distribution in the $1/\nu$ regime is therefore found by integrating the expression
\begin{eqnarray}\label{eq:DKE_1overnu}
  \sum_\sigma
\frac{Z e\Psi}{m}\,
\int_{l_{b_{10}}}^{l_{b_{20}}}
|v_{||}^{(0)}|^{-1}
{C_{ii}^{\ell (0)}\left[g_{1/\nu}^{(1)}\right]}\dd l
=
\partial_\alpha J^{(1)} \Upsilon F_{M}.
\end{eqnarray}
Below, we will find useful to distinguish between $g_{1/\nu}^{(1)}[J_B^{(1)}]$ and $g_{1/\nu}^{(1)}[J_\varphi^{(1)}]$, where
\begin{eqnarray}\label{eq:DKE_1overnu_B1}  \sum_\sigma
\frac{Z e\Psi}{m}\,
\int_{l_{b_{10}}}^{l_{b_{20}}}
|v_{||}^{(0)}|^{-1}
{C_{ii}^{\ell (0)}\left[g_{1/\nu}^{(1)}[J_B^{(1)}]\right]}\dd l
=
\partial_\alpha J_B^{(1)} \Upsilon F_{M}
\end{eqnarray}
and
\begin{eqnarray}\label{eq:DKE_1overnu_varphi1}  \sum_\sigma
\frac{Z e\Psi}{m}\,
\int_{l_{b_{10}}}^{l_{b_{20}}}
|v_{||}^{(0)}|^{-1}
{C_{ii}^{\ell (0)}\left[g_{1/\nu}^{(1)}[J_\varphi^{(1)}]\right]}\dd l
=
\partial_\alpha J_\varphi^{(1)} \Upsilon F_{M},
\end{eqnarray}
and therefore $g_{1/\nu}^{(1)} = g_{1/\nu}^{(1)}[J_B^{(1)}] + g_{1/\nu}^{(1)}[J_\varphi^{(1)}]$. We note that
\begin{equation}\label{eq:size_g1overnu_B1}
g_{1/\nu}^{(1)}[J_B^{(1)}] \sim \frac{\rho_*}{\nu_*} \frac{n}{v_t^3}
\end{equation}
and
\begin{equation}\label{eq:size_g1overnu_varphi1}
g_{1/\nu}^{(1)}[J_\varphi^{(1)}] \sim \frac{\rho_*}{\nu_*} \frac{e\varphi^{(1)}}{T}\frac{n}{v_t^3}.
\end{equation}

The equation for $\varphi^{(1)}$ reads (recall \eq{eq:quasineutrality_equation})
\begin{eqnarray}\label{eq:QN_1overnu_preliminar}
&&\hspace{-1cm}\varphi^{(1)} =\left(\frac{Z}{T}+\frac{1}{T_e}\right)^{-1}
\frac{2\pi B_0}{en} 
\int_0^\infty\dd v v^3 \int_{B^{-1}_{0,{\rm max}}}^{B_0^{-1}}\dd\lambda
|v_{||}^{(0)}|^{-1}
\left(
 g_{1/\nu}^{(1)}[J_\varphi^{(1)}] + g_{1/\nu}^{(1)}[J_B^{(1)}]
\right).
\end{eqnarray}
Using \eq{eq:size_g1overnu_varphi1} and $\rho_* \ll \nu_*$, and comparing the size of the left-hand side of \eq{eq:QN_1overnu_preliminar} with the size of the term on the right-hand side containing $g_{1/\nu}^{(1)}[J_\varphi^{(1)}]$, we conclude that the latter can be dropped. Hence, $\varphi^{(1)}$ is calculated as
\begin{eqnarray}\label{eq:QN_1overnu}
\varphi^{(1)} =\left(\frac{Z}{T}+\frac{1}{T_e}\right)^{-1}
\frac{2\pi B_0}{en} 
\int_0^\infty\dd v v^3 \int_{B^{-1}_{0,{\rm max}}}^{B_0^{-1}}\dd\lambda
|v_{||}^{(0)}|^{-1}
g_{1/\nu}^{(1)}[J_B^{(1)}]
.
\end{eqnarray}
Then, in the $1/\nu$ regime (see \eq{eq:size_g1overnu_B1} and \eq{eq:size_g1overnu_varphi1}),
\begin{equation}\label{eq:g1_1overnu_tightepsilon}
g^{(1)} \sim \frac{\rho_*}{\nu_*} \frac{n}{v_t^3}
\end{equation}
and
\begin{equation}\label{eq:scaling_varphi1_1overnu_regime}
\varphi^{(1)} \sim \frac{\rho_*}{\nu_*}\frac{T}{e}.
\end{equation}

If $B_0$ and $B_1$ are stellarator symmetric, $J_B^{(1)}$ is symmetric and, consequently, $\partial_\alpha J_B^{(1)}$ is antisymmetric. Since the collision operator does not change the symmetry, the solution $g_{1/\nu}^{(1)}$ of \eq{eq:DKE_1overnu} is antisymmetric. Finally, $\varphi^{(1)}$, obtained from \eq{eq:QN_1overnu}, is stellarator antisymmetric.

\subsubsection{The $1/\nu$ regime in a large aspect ratio stellarator}
\label{sec:varphi1_1overnu_regime_numerical}

As indicated by the end of subsection \ref{sec:large_aspect_ratio}, when $\epsilon$ is small the low collisionality condition reads $\nu_* \ll \epsilon^{3/2}$. How small $\nu_*$ can be before the tangential drifts start to count depends on their size, as will be discussed mainly in subsections \ref{sec:sqrtnu_large_Er} and \ref{sec:sqrtnu_small_Er}.

Let us refine the estimates \eq{eq:g1_1overnu_tightepsilon} and \eq{eq:scaling_varphi1_1overnu_regime} for the distribution function and the electrostatic potential when $\epsilon\ll 1$. The drift-kinetic equation for $\epsilon\ll 1$ is
\begin{eqnarray}\label{eq:DKE_1overnu_smallepsilon}
\frac{2 Z e\Psi}{m B_{00}^2}\,
\partial_\lambda
\left[
\left(\int_{l_{b_{10}}}^{l_{b_{20}}}
\sqrt{1-\lambda B_0} \, \dd l
\right)
\partial_\lambda g_{1/\nu}^{(1)}
\right]
=\frac{v}{\nu_\lambda}
\partial_\alpha J^{(1)} \Upsilon F_{M}.
\end{eqnarray}

Employing the scalings of subsection \ref{sec:large_aspect_ratio}, it is easy to infer that
\begin{equation}\label{eq:size_g1overnu_B1_smallepsilon}
g_{1/\nu}^{(1)}[J_B^{(1)}] \sim \frac{\rho_*}{\nu_*} \frac{n}{v_t^3}
\end{equation}
and
\begin{equation}\label{eq:size_g1overnu_varphi1_smallepsilon}
g_{1/\nu}^{(1)}[J_\varphi^{(1)}] \sim \epsilon^{-1/2}\frac{\rho_*}{\nu_*} \frac{J_\varphi^{(1)}}{v_t R_0}\frac{n}{v_t^3}.
\end{equation}

\begin{figure}
\centering
\includegraphics[width=0.8\textwidth]{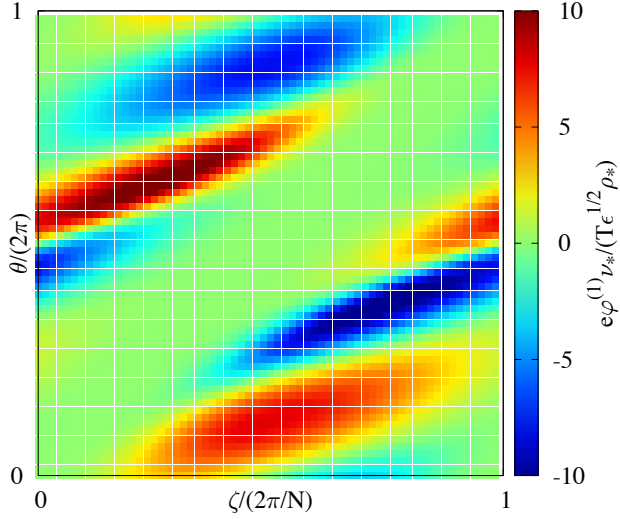}
\caption{Contour plot of $\varphi^{(1)}$ for $\epsilon = 1.1 \times 10^{-4}$, $\nu_*= 10^{-7}$ and $\varphi'_0 = 0$.}
\label{fig:contour_plot_usual_1overnu_regime}
\end{figure}

Next, we apply the $\epsilon\ll 1$ expansion to \eq{eq:QN_1overnu_preliminar}. We must treat two cases separately. If
\begin{equation}\label{eq:1overnu_standard}
\epsilon^{-1/2}\rho_* \ll \nu_* \ll \epsilon^{3/2}, 
\end{equation}
then the first term on the right side of \eq{eq:QN_1overnu_preliminar} is negligible with respect to the left-hand side. Using that $\Delta_{{\rm trapped}}\sim \epsilon B_{00}^{-1}$, we find
\begin{equation}\label{eq:scaling_varphi1_1overnu_regime_smallepsilon}
\varphi^{(1)} \sim \epsilon^{1/2}\frac{\rho_*}{\nu_*}\frac{T}{e}.
\end{equation}
And noting that $g_{1/\nu}^{(1)}[J_\varphi^{(1)}] \ll g_{1/\nu}^{(1)}[J_B^{(1)}]$,
\begin{equation}\label{eq:scaling_g1_1overnu_regime_smallepsilon}
g^{(1)} \sim \frac{\rho_*}{\nu_*} \frac{n}{v_t^3}.
\end{equation}
Then, both $\varphi^{(1)}$ and $g^{(1)}$ scale as $\nu_*^{-1}$. The fact that $\varphi^{(1)}$ is antisymmetric is also clear. In figure \ref{fig:contour_plot_usual_1overnu_regime}, a contour plot of $\varphi^{(1)}$ on the magnetic surface is provided for a set of values such that \eq{eq:1overnu_standard} holds.

However, if $\nu_* \sim \epsilon^{-1/2}\rho_*$ or smaller, the first term on the right-hand side of \eq{eq:QN_1overnu_preliminar} can become comparable to the left-hand side. Let us study the asymptotic regime
\begin{equation}\label{eq:def_intermediate_regime}
\nu_* \ll \epsilon^{-1/2}\rho_*.
\end{equation}

From \eq{eq:QN_1overnu_preliminar} we learn that $\varphi^{(1)}$ depends on $l$ only through $B_0$. This will be important in a moment. When \eq{eq:def_intermediate_regime} holds, the left-hand side of \eq{eq:QN_1overnu_preliminar} is small compared to the term on the right-hand side that involves $\varphi^{(1)}$. Then, we can write 
\begin{eqnarray}\label{eq:QN_1overnu_subregime}
&& 
\int_0^\infty\dd v v^2 \int_{B^{-1}_{0,{\rm max}}}^{B_0^{-1}}\dd\lambda
\frac{1}{\sqrt{1-\lambda B_0}}\,
 g_{1/\nu}^{(1)} = 0,
\end{eqnarray}
where $g_{1/\nu}^{(1)}$ is the solution of \eq{eq:DKE_1overnu_smallepsilon}. For the argument that follows, we find it convenient to write $J^{(1)}$ as
\begin{equation}\label{eq:J1_coor_B}
J^{(1)}
=
 -  \int_{B_{0,{\rm min}}}^{B_0}
\frac{
 G
}
{\sqrt{1-\lambda {\hat B}}}\, \dd {\hat B}.
\end{equation}
Here, $B_{0,{\rm min}}$ is the minimum of $B_0$,
\begin{equation}\label{eq:aux_function_G}
G(r,\alpha, {\hat B}, \beta, v) := \sum_{\beta = -1}^1 \left(\frac{B_1 v}{B_{00}} + \frac{2Ze\varphi^{(1)}}{mv}\right)
\frac{1}{\left|
\partial_l B_0
\right|}
\end{equation}
and we have used the magnitude of $B_0$ as the integration variable in \eq{eq:J1_coor_B}. Since there are two values\footnote{For simplicity in the presentation, we assume that the omnigeneous wells are such that exactly two points in the well take each value $B_0 = {\hat B}$ (except for the value $B_{0,{\rm min}}$, which is reached at a single point). There exist omnigeneous magnetic fields with more complicated wells \citep{Parra2015}, and in those cases the discrete coordinate $\beta$ would need to label more than two branches.} of $l$ corresponding to each value $B_0 = {\hat B}$, we introduce the discrete coordinate $\beta = \pm 1$. The value $\beta=-1$ labels the branch of the magnetic well where $\partial_l B_0 < 0$ and the value $\beta=1$ labels the branch where $\partial_l B_0 \ge 0$. On the right-hand side of \eq{eq:aux_function_G}, every function is understood to be expressed using ${\hat B}$ and $\beta$ instead of $l$. In particular, we note that, in general, $|\partial_l B_0|({\hat B}, \beta) \ne |\partial_l B_0|({\hat B}, -\beta)$.

It is clear that
\begin{eqnarray}\label{eq:QN_1overnu_subregime_2}
&& 
\int_0^\infty v^2 g_{1/\nu}^{(1)} \dd v = 0
\end{eqnarray}
gives the solution of \eq{eq:QN_1overnu_subregime} and that, by imposing \eq{eq:QN_1overnu_subregime_2}, we obtain a relation between the orbit integrals of $B_1$ and $\varphi^{(1)}$. The orbit integral of a function can be interpreted as a transformation that replaces the coordinate $l$ by the coordinate $\lambda$. This can be viewed as an Abel transform, and an explicit inversion formula is given in Appendix \ref{sec:Abel_transform}. In principle, the inversion cannot distinguish between two points of the magnetic well with the same value of $B_0$ (this is quite intuitive when we use the magnitude of $B_0$ as the integration variable, as we have done in \eq{eq:J1_coor_B}). Carrying out the integral in $v$ on the left-hand side of \eq{eq:QN_1overnu_subregime_2}, we arrive at
\begin{eqnarray}
&&\hspace{-1cm}
\sum_{\beta = -1}^1
\frac{Ze\varphi^{(1)}}{T}
\frac{1}{| \partial_l B_0 |}
=
-2.83 \,
\frac{n'/n + Ze\varphi'_0/T + 2.33 T'/T}{n'/n + Ze\varphi'_0 / T + 1.33 T'/T}
\sum_{\beta = -1}^1
\frac{B_1}{B_{00}}
\frac{1}{|\partial_l B_0 |}
.
\end{eqnarray}
Recalling that $\varphi^{(1)}$ depends on $l$ only through $B_0$ (see equation \eq{eq:QN_1overnu_preliminar}), we finally obtain the explicit solution
\begin{eqnarray}\label{eq:explicit_solution_varphi1_subregime1overnu}
&&\hspace{-1cm}
\frac{Ze\varphi^{(1)}}{T}
=
-2.83 \,
\frac{n'/n + Ze\varphi'_0/T + 2.33 T'/T}{n'/n + Ze\varphi'_0 / T + 1.33 T'/T}
\,
\frac{
\sum_{\beta = -1}^1
(B_1/B_{00})|\partial_l B_0 |^{-1}
}
{
\sum_{\beta = -1}^1
|\partial_l B_0|^{-1}
}
.
\end{eqnarray}

From \eq{eq:explicit_solution_varphi1_subregime1overnu} we infer that in this regime $\varphi^{(1)}$ is stellarator symmetric and that its size, in general, is
\begin{equation}\label{eq:size_varphi1_new1overnu}
\varphi^{(1)} \sim \epsilon \frac{T}{e}.
\end{equation}
We note that for particular combinations of the profile gradients, $\varphi^{(1)}$ can get larger because the denominator of \eq{eq:explicit_solution_varphi1_subregime1overnu} can become small. This will be studied elsewhere.

\begin{figure}
\centering
\includegraphics[width=0.8\textwidth]{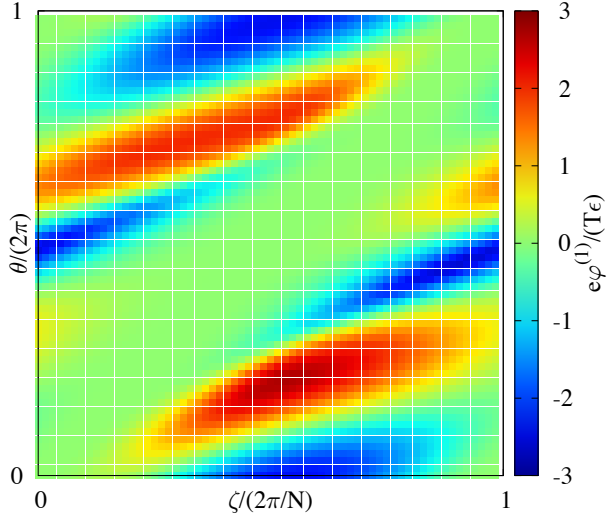}
\caption{Contour plot of $\varphi^{(1)}$ for $\epsilon = 1.1\times 10^{-4}$, $\nu_*=5.8\times 10^{-11}$ and $\varphi'_0 = 0$.}
\label{fig:contour_plot_intermediate_regime}
\end{figure}

In figure \ref{fig:contour_plot_intermediate_regime}, a contour plot of $\varphi^{(1)}$ on the magnetic surface is provided for a set of values that satisfy \eq{eq:def_intermediate_regime}.

Finally, we have
(see \eq{eq:size_g1overnu_B1_smallepsilon} and \eq{eq:size_g1overnu_varphi1_smallepsilon})
\begin{equation}\label{eq:size_g_intermediate_regime}
g_{1/\nu}^{(1)} \sim g_{1/\nu}^{(1)}[J_\varphi^{(1)}] \sim g_{1/\nu}^{(1)}[J_B^{(1)}] \sim \frac{\rho_*}{\nu_*} \frac{n}{v_t^3}.
\end{equation}

\begin{figure}
\centering
\includegraphics[width=0.8\textwidth]{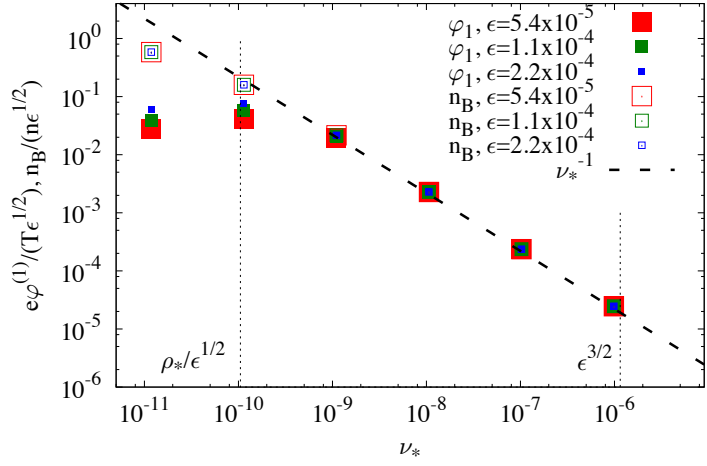}
\caption{Scaling of $\varphi^{(1)}$ and $n_B$ (evaluated at the point of the flux surface where they reach their maximum values) with $\nu_*$ and $\epsilon$. Here, $\varphi'_0 = 0$. The regimes  \eq{eq:1overnu_standard} and \eq{eq:def_intermediate_regime} are confirmed numerically. The central value $\epsilon = 1.1\times 10^{-4}$ has been used to draw the thin dotted lines indicating where the collisionality values $\rho_*/\epsilon^{1/2}$ and $\epsilon^{3/2}$ are located.}
\label{fig:scaling_1overnu_and_intermediate_regimes}
\end{figure}

In figure \ref{fig:scaling_1overnu_and_intermediate_regimes}, and for $\varphi'_0 = 0$, we illustrate the scaling of $\varphi^{(1)}$ and of
\begin{equation}\label{eq:def_nB}
n_B:=\int g_{1/\nu}^{(1)}[J_B^{(1)}]\dd^3 v \sim  \frac{\rho_*}{\nu_*} n
\end{equation}
in a range of collisionality values such that both regimes, \eq{eq:1overnu_standard} and \eq{eq:def_intermediate_regime}, are observed. Whereas $\varphi^{(1)}$ and $n_B$ have the same scaling with $\nu_*$ in the regime \eq{eq:1overnu_standard}, they scale differently in the regime \eq{eq:def_intermediate_regime}.

\subsection{The $\sqrt{\nu}$ regime}
\label{sec:varphi1_sqrtnu_regime}

We turn to discuss collisionality regimes with $\nu_* \ll \rho_* \ll 1$. In this situation, as explained at length in \citet{CalvoPPCF2017}, the behavior of both the radial neoclassical fluxes and $\varphi^{(1)}$ depends on the zeroes of $\partial_\rcoor J^{(0)}$.

When $\nu_* \ll \rho_*$, the collision term in \eq{eq:DKEtrapped} is negligible compared to the first term on the left-hand side,
\begin{equation}\label{eq:collisions_negligible}
\left|
 \sum_\sigma
\frac{Z e\Psi}{m}\,
\int_{l_{b_{10}}}^{l_{b_{20}}}\frac{\dd l}{|v_{||}^{(0)}|}
{C_{ii}^{\ell (0)}[g^{(1)}]}
\right|
\ll
\left|
\partial_\rcoor J^{(0)} \partial_\alpha  g^{(1)}
\right|
.
\end{equation}
Hence, to lowest order in an expansion in $\nu_*/\rho_*$,
\begin{equation}\label{eq:expansion_sqrtnu}
g^{(1)} = g_0 + \dots,
\end{equation}
where $g_0$ is the solution of
\begin{eqnarray}\label{eq:DKEg0}
-\partial_\rcoor J^{(0)}
\partial_\alpha  g_0
  + \partial_\alpha J^{(1)} \Upsilon F_{M}
 = 0.
\end{eqnarray}
This equation can be readily solved, obtaining an explicit expression for $g_0$,
\begin{eqnarray}\label{eq:g0}
g_0
=
\frac{1}{\partial_\rcoor J^{(0)}}\left(
J^{(1)}-\frac{1}{2\pi}\int_0^{2\pi}J^{(1)}\dd\alpha
\right)\Upsilon F_{M},
\end{eqnarray}
where we have chosen $\int_0^{2\pi} g_0 \dd\alpha = 0$ to fix the integration constant. From \eq{eq:g0}, it is clear that the solution $g_0$ is not valid at phase space points where $\partial_\rcoor J^{(0)}$ vanishes, and we devote subsection \ref{sec:varphi1_sb-p_regime} to those points. However, if $\partial_\rcoor J^{(0)}$ does not vanish at any point of phase space (or if it vanishes only for values of the velocity $v\gg v_t$, that are irrelevant because their effect is supressed by the smallness of the Maxwellian distribution), \eq{eq:g0} is enough to correctly determine $\varphi^{(1)}$ via the quasineutrality equation
\begin{eqnarray}\label{eq:QN_1sqrtnu}
\varphi^{(1)} =\left(\frac{Z}{T}+\frac{1}{T_e}\right)^{-1}
\frac{2\pi B_0}{en} 
\int_0^\infty\dd v v^3 \int_{B^{-1}_{0,{\rm max}}}^{B_0^{-1}}\dd\lambda
|v_{||}^{(0)}|^{-1}
g_0.
\end{eqnarray}
In this case, $\varphi^{(1)}$ is stellarator symmetric (due to the fact that the expression \eq{eq:g0} for $g_0$ does not involve derivatives of $B$ along the flux surface) and does not scale with collisionality,
\begin{equation}
\varphi^{(1)} \sim \frac{T}{e}.
\end{equation}

As proven in \citet{CalvoPPCF2017}, $g_0$ does not contribute to the radial neoclassical fluxes. They are determined by corrections to $g_0$ in \eq{eq:expansion_sqrtnu} that are localized in a region of phase space that has a typical size $\Delta_{\sqrt{\nu}}\sim \sqrt{\nu_*/\rho_*}$ in the coordinate $\lambda$. This small layer produces the scaling with collisionality that justifies the name of this regime, the $\sqrt{\nu}$ regime.

If the aspect ratio is large, it is interesting to distinguish two cases (mostly in connection to the solutions of the quasineutrality equation), to which we devote subsections \ref{sec:sqrtnu_large_Er} and \ref{sec:sqrtnu_small_Er}.

\subsubsection{Large aspect ratio and large tangential drift, $\partial_\rcoor J^{(0)} \sim \epsilon^{-3/2} v_t R_0$}
\label{sec:sqrtnu_large_Er}

If $\partial_\rcoor J^{(0)} \sim \epsilon^{-3/2} v_t R_0$, the inequality \eq{eq:collisions_negligible} holds provided that $\nu_* \ll \epsilon^{-1}\rho_*$. Hence, the $1/\nu$ regime takes place in the range
\begin{equation}\label{eq:def_1overnu_large_Er}
\epsilon^{-1} \rho_* \ll \nu_* \ll \epsilon^{3/2}.
\end{equation}
In particular, the regime defined by \eq{eq:def_intermediate_regime} is not observed because the tangential drift starts to count for collisionalities much larger than $\epsilon^{-1/2}\rho_*$. It can be observed, however, when the tangential drift is smaller (see subsection \ref{sec:sqrtnu_small_Er}).

As pointed out in subsection \ref{sec:large_aspect_ratio}, $\partial_\rcoor J^{(0)} \sim \epsilon^{-3/2} v_t R_0$  corresponds to $\varphi'_0 \sim \epsilon^{-1} T / (e R_0)$, which is the standard size of the radial electric field in a large aspect ratio stellarator. The derivation of the radial fluxes in this regime can be found, for example, in \citet{Galeev1979, Ho1987}.

 Let us start the discussion about $\varphi^{(1)}$ by writing \eq{eq:QN_1sqrtnu} more explicitly. Inserting in \eq{eq:QN_1sqrtnu} the expression for $g_0$ (see \eq{eq:g0}) and rearranging a bit, we get
\begin{eqnarray}\label{eq:QNsqrtnu_rearranged}
\varphi^{(1)} - 
\frac{2\pi}{en} 
\left(\frac{Z}{T}+\frac{1}{T_e}\right)^{-1}
\int_0^\infty\dd v \int_{B^{-1}_{0,{\rm max}}}^{B_0^{-1}}\dd\lambda
\frac{v^3 B_0}{|v_{||}^{(0)}|}
\frac{\Upsilon F_{M}}{\partial_\rcoor J^{(0)}}
{\widetilde J_\varphi^{(1)}}
\nonumber\\
\hspace{0.5cm}
=
\frac{2\pi}{en} \left(\frac{Z}{T}+\frac{1}{T_e}\right)^{-1}
\int_0^\infty\dd v \int_{B^{-1}_{0,{\rm max}}}^{B_0^{-1}}\dd\lambda
\frac{v^3 B_0}{|v_{||}^{(0)}|}
\frac{\Upsilon F_{M}}{\partial_\rcoor J^{(0)}}
{\widetilde J_B^{(1)}}
,
\end{eqnarray}
where
\begin{equation}\label{eq:JB1tilde}
{\widetilde J_B^{(1)}} = J_B^{(1)}-\frac{1}{2\pi}\int_0^{2\pi}J_B^{(1)}\dd\alpha
\end{equation}
and
\begin{equation}\label{eq:J1_varphi_tilde}
{\widetilde J_\varphi^{(1)}} = J_\varphi^{(1)}-\frac{1}{2\pi}\int_0^{2\pi}J_\varphi^{(1)}\dd\alpha.
\end{equation}
In \eq{eq:J1_varphi_tilde} we have emphasized that only the component of $J_\varphi^{(1)}$ that fluctuates in $\alpha$ enters \eq{eq:QNsqrtnu_rearranged} although, strictly speaking, it is superfluous due to \eq{eq:property_J1_varphi}.

\begin{figure}
\centering
\includegraphics[width=0.8\textwidth]{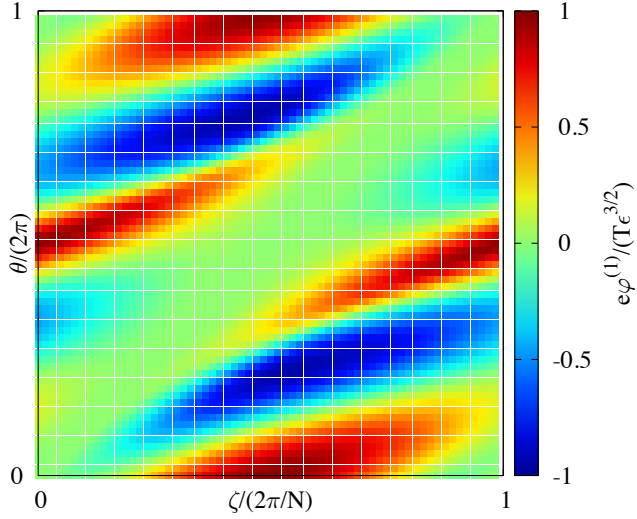}
\caption{Contour plot of $\varphi^{(1)}$ for $\epsilon = 1.1\times 10^{-4}$, $\nu_*= 1.1\times 10^{-10}$ and $\varphi'_0 = 2 \epsilon^{-1} T/(eR_0)$.}
\label{fig:contour_plot_sqrtnu_regime}
\end{figure}

\begin{figure}
\centering
\includegraphics[width=0.8\textwidth]{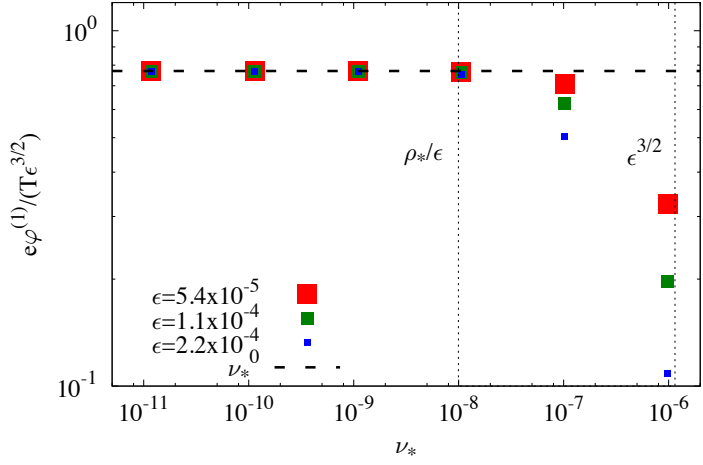}
\caption{Scaling of $\varphi^{(1)}$ with $\nu_*$ and $\epsilon$ (evaluated at the point of the flux surface where it reaches its maximum value) in a range of values of the collisionality where \eq{eq:scaling_varphi1_sqrtnu_regime} is observed. Here, $\varphi'_0 = 2 \epsilon^{-1} T/(eR_0)$.}
\label{fig:scaling_sqrtnu_regime}
\end{figure}

Recalling the estimates in subsection \ref{sec:large_aspect_ratio}, it is straightforward to see that the second term on the left-hand side of \eq{eq:QNsqrtnu_rearranged} is smaller than the first one by a factor $\epsilon^{1/2}$. Hence, the size of $\varphi^{(1)}$ is determined by the size of the right-hand side of \eq{eq:QNsqrtnu_rearranged}. That is,
\begin{equation}\label{eq:scaling_varphi1_sqrtnu_regime}
\varphi^{(1)} \sim \epsilon^{3/2} \, \frac{T}{e}\,.
\end{equation}
This scaling is consistent with \eq{eq:scaling_varphi1_1overnu_regime_smallepsilon} when $\nu_*\sim \epsilon^{-1}\rho_*$. In figure \ref{fig:contour_plot_sqrtnu_regime}, a contour plot of $\varphi^{(1)}$ corresponding to  $\partial_\rcoor J^{(0)} \sim \epsilon^{-3/2} v_t R_0$ and $\nu_* \ll \epsilon^{-1}\rho_*$ is shown. In figure \ref{fig:scaling_sqrtnu_regime}, the collisionality-independent scaling \eq{eq:scaling_varphi1_sqrtnu_regime} for $\varphi^{(1)}$ is numerically checked.

\subsubsection{Large aspect ratio and small tangential drift, $\partial_\rcoor J^{(0)} \sim \epsilon^{-1/2} v_t R_0$}
\label{sec:sqrtnu_small_Er}

If $\partial_\rcoor J^{(0)} \sim \epsilon^{-1/2} v_t R_0$, \eq{eq:collisions_negligible} is satisfied when $\nu_* \ll \rho_*$. Therefore, for such typical size of the tangential drifts, the $1/\nu$ regime happens for collisionality values
\begin{equation}\label{eq:def_1overnu_small_Er}
\rho_* \ll \nu_* \ll \epsilon^{3/2}.
\end{equation}
To be precise, in the range $\rho_* \ll \nu_* \ll \epsilon^{-1/2}\rho_*$, the subregime \eq{eq:def_intermediate_regime} of the $1/\nu$ regime will be observed.

When $\partial_\rcoor J^{(0)} \sim \epsilon^{-1/2} v_t R_0$, the left-hand side of \eq{eq:QN_1sqrtnu} can be neglected compared to the term on the right-hand side containing $\varphi^{(1)}$, so that we have
\begin{eqnarray}\label{eq:QN_1sqrtnu_small_tang_drift} 
\int_0^\infty\dd v v^2 \int_{B^{-1}_{0,{\rm max}}}^{B_0^{-1}}\dd\lambda
\frac{g_0}{\sqrt{1-\lambda B_0}}
 = 0,
\end{eqnarray}
which can be solved by finding the solution of
\begin{eqnarray}\label{eq:QN_1sqrtnu_small_tang_drift_2} 
\int_0^\infty\dd v v^2
g_0
 = 0.
\end{eqnarray}
One could formally proceed as done after equation \eq{eq:def_intermediate_regime}, but it seems difficult to find a completely explicit solution for $\varphi^{(1)}$ due to the relatively complicated dependence of $\partial_r J^{(0)}$ on $v$ and $\lambda$. In any case, it is clear that imposing \eq{eq:QN_1sqrtnu_small_tang_drift_2} implies
\begin{equation}
J_\varphi^{(1)} \sim \epsilon^{1/2} v_t R_0
\end{equation}
and, therefore,
 \begin{equation}\label{eq:size_varphi_sqrtnu}
\varphi^{(1)} \sim \epsilon \frac{T}{e}.
\end{equation}
In addition, from \eq{eq:QN_1sqrtnu_small_tang_drift_2} we can infer that $\varphi^{(1)}$ is stellarator symmetric.

Note that for $\nu_*\sim \rho_*$, expression \eq{eq:size_varphi_sqrtnu} matches \eq{eq:size_varphi1_new1overnu}.

\subsection{Superbanana-plateau regime}
\label{sec:varphi1_sb-p_regime}

In this subsection, we treat cases with $\nu_* \ll \rho_* \ll 1$ and such that $\partial_\rcoor J^{(0)}$ vanishes for $v\lesssim v_t$. Subsections \ref{sec:varphi1_1overnu_regime} and \ref{sec:varphi1_sqrtnu_regime}, although brief, were self-contained to a large extent. In this subsection we will refer the reader to \citet{CalvoPPCF2017} more frequently for the derivation of particularly technical results.

The zeroes of $\partial_\rcoor J^{(0)}$ correspond to `resonant' points of phase space where the orbit-averaged tangential component of the magnetic and $E\times B$ drifts cancel each other.

Recall equation \eq{eq:dJ0/dpsi}. The condition $\partial_\rcoor J^{(0)} = 0$ can be compactly expressed as
\begin{equation}\label{eq:resonance_condition}
\lambda \overline{\partial_\rcoor B_0}(\rcoor,\lambda)
=
-\frac{2Z e \varphi'_0(\rcoor)} 
{m v^2},
\end{equation}
where
\begin{equation}
\overline{(\cdot)} = \frac{1}{\tau_b^{(0)}}\sum_\sigma
\int_{l_{b_{10}}}^{l_{b_{20}}}(\cdot) |v_{||}^{(0)}|^{-1} \dd l
\end{equation}
denotes orbit average and
\begin{equation}
\tau_b^{(0)}(\rcoor,v,\lambda) = 2
\int_{l_{b_{10}}}^{l_{b_{20}}} |v_{||}^{(0)}|^{-1} \dd l
\end{equation}
is the time that it takes for the particle to complete the orbit in the magnetic field $B_0$. A necessary (but not sufficient) condition for this equation to have solutions for $v\lesssim v_t$ is
\begin{equation}\label{eq:sizeEr_for_resonance}
\frac{Ze|\varphi'_0|}{T} \lesssim \frac{1}{R_0}.
\end{equation}
That is, the effect of the resonances is expected to be important for relatively small values of the radial electric field. When $\nu_* \ll \rho_* \ll 1$ and $Ze|\varphi'_0|/T \gg R_0^{-1}$, the treatment of subsection \eq{sec:varphi1_sqrtnu_regime} applies.

From here on, let us assume $\nu_* \ll \rho_* \ll 1$ and condition \eq{eq:sizeEr_for_resonance} in the rest of this subsection, and let us also assume that, for $v\lesssim v_t$, there exists one solution in $\lambda$ of the resonance condition \eq{eq:resonance_condition}, that we denote by $\lambda_r(\rcoor,v,\varphi'_0)$. In general, $\lambda_r$ will be defined only for some values of $\rcoor$, $v$ and $\varphi'_0$. For fixed $\rcoor$ and $\varphi'_0$, we denote by $I$ the interval in $v$ for which $\lambda_r$ is defined.

We can expand the drift-kinetic equation \eq{eq:DKEtrapped} around the position of the resonance. Defining $\xi = \lambda - \lambda_r(r,v,\varphi'_0)$ and writing the drift-kinetic equation in coordinates $v$ and $\xi$, we have
\begin{eqnarray}\label{eq:DKEresonantlayer}
\partial_\lambda \partial_\rcoor J_r^{(0)}
\xi
\partial_\alpha  g_{\rm rl}
 +
k \partial_\xi^2 g_{\rm rl}
= \left(
\partial_\alpha J_{B,r}^{(1)}
+
\partial_\alpha \widehat{J_\varphi^{(1)}}
\right)
\Upsilon F_M,
\end{eqnarray}
where $k(r,v)$ is given by
\begin{equation}\label{eq:def_k}
k = \frac{2Ze\Psi}{mv}
\Bigg[
\nu_\lambda \lambda_r \int_{l_{b_{10}}}^{l_{b_{20}}} B_0^{-1}\sqrt{1-\lambda_r B_0}\, \dd l
+
\frac{\nu_v v^2}{2} (\partial_v \lambda_r)^2 \int_{l_{b_{10}}}^{l_{b_{20}}} \frac{1}{\sqrt{1-\lambda_r B_0}}\, \dd l
\Bigg],
\end{equation}
subindices $r$ indicate that the corresponding quantity is evaluated at $\lambda = \lambda_r(r,v,\varphi'_0)$
and $\widehat{J_\varphi^{(1)}}$ is an approximation of $J_\varphi^{(1)}$ around $\lambda_r$. In particular, $\widehat{J_\varphi^{(1)}}$ is a function of $v$ and $\xi$. We will see shortly why, in general, we cannot simply evaluate $J_\varphi^{(1)}$ at $\lambda_r$. The second term in \eq{eq:def_k} comes from the energy diffusion piece of the collision operator, which must be included because the coordinate across the layer (the coordinate in which $g_{\rm rl}$ varies fast), $\xi = \lambda - \lambda_r$, mixes $v$ and $\lambda$. The collision frequency $\nu_v$ is given by
\begin{equation}
\nu_v(v)
=
\frac{3\sqrt{\pi}\, \chi\left(v\sqrt{m/(2T)}\right)}{2\tau_{ii}\left(v\sqrt{m/(2T)}\right)^3}\, .
\end{equation}
In \citet{CalvoPPCF2017}, the energy diffusion piece of the collision operator was incorrectly neglected. Note that none of the scalings derived in  Calvo et al (2017) are affected by the inclusion of energy diffusion.

Using that $\partial_\lambda \partial_\rcoor J_r^{(0)} = O(B_0 v_t)$ and that $k(\rcoor,v) = O(\nu_\lambda B_0^{-2} R_{0} \rho_*^{-1})$, and balancing the two terms on the left-hand side of \eq{eq:DKEresonantlayer}, we deduce that $g_{\rm rl}$ is localized in a layer whose size in the coordinate $\xi$ is
 \begin{equation}\label{eq:size_sb-p_layer}
 B_0 \Delta_{{\rm sb-p}} \sim (\nu_* / \rho_*)^{1/3}.
 \end{equation}
Then, balancing the left- and right-hand sides of \eq{eq:DKEresonantlayer}, we find that
\begin{equation}
g_{\rm rl}
\sim  (B_0 \Delta_{{\rm sb-p}})^{-1} F_{M}.
\end{equation}
This resonant layer, via $g_{\rm rl}$, is responsible for superbanana-plateau transport.
 
The solution outside the resonant layer is given by \eq{eq:g0}. Since $g_0$ diverges at $\lambda_r$, we remove the divergence so that we obtain a function $g_0^{\rm out}$ which is finite everywhere and asymptotically coincides with $g_0$ far from the resonant layer. Namely,
\begin{eqnarray}
\hspace{-0.65 cm} g_0^{\rm out}
=
g_0
-
\frac{1}{(\lambda - \lambda_r)\partial_\lambda\partial_\rcoor J_r^{(0)}}
\Bigg(
{\widetilde J_{B,r}^{(1)}}
+
\widehat{J_\varphi^{(1)}}-\frac{1}{2\pi}\int_0^{2\pi}\widehat{J_\varphi^{(1)}}
\dd\alpha
\Bigg)\Upsilon F_M,
\end{eqnarray}
where ${\widetilde J_{B,r}^{(1)}}$ is the quantity defined in \eq{eq:JB1tilde} evaluated at $\lambda_r$.

\begin{figure}
\centering
\includegraphics[width=0.6\textwidth]{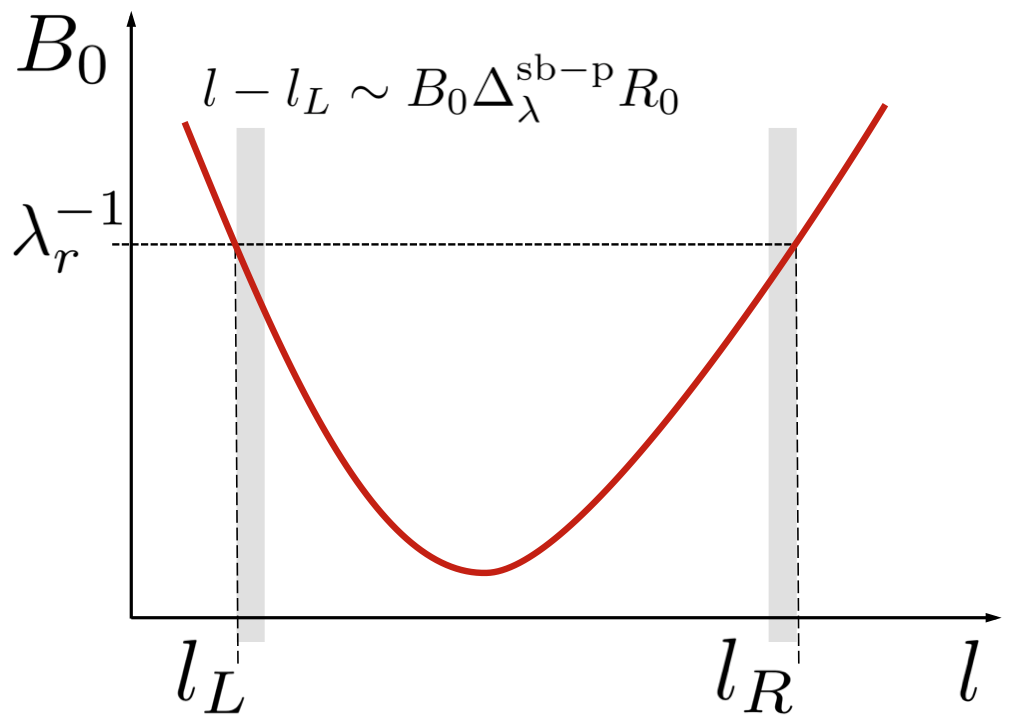}
\caption{If \eq{eq:sb-p_small_Er} is satisfied, the resonant value $\lambda_r$ is approximately independent of $v$. The accumulated effect of the resonance for all possible values of $v$ produces a larger $\varphi^{(1)}$ in a neighborhood of the bounce points of the resonant trajectory, $l_L$ and $l_R$.}
\label{fig:resonant_trajectory}
\end{figure}

In principle, both pieces, $g_0^{\rm out}$ and $g_{\rm rl}$, contribute equally to the quasineutrality equation (although only $g_{\rm rl}$ contributes to radial transport). Hence, one needs to solve
\begin{eqnarray}\label{eq:QNorderepsilon2}
\hspace{-0.7 cm}
\varphi^{(1)} =
  \frac{2\pi B_0}{en}\left(\frac{Z}{T}+\frac{1}{T_e}\right)^{-1} \left[
  \int_0^\infty\dd v
  \int_{B^{-1}_{0,{\rm max}}}^{B^{-1}}\dd\lambda
  \frac{v^3 g_0^{\rm out}}{|v_{||}^{(0)}|}
  +
  \int_I \dd v
  \int_{K\Delta_{{\rm sb-p}}}\dd\lambda
  \frac{v^3 g_{\rm rl}}{|v_{||}^{(0)}|}
  \right],
\end{eqnarray}
with $K \gg 1$. An explicit, asymptotically correct expression for the quasineutrality equation is given by (see \citet{CalvoPPCF2017} for the details)
\begin{eqnarray}\label{eq:QN_sb-p_explicit}
&&
\varphi^{(1)} =
  \frac{2\pi B_0}{en} \left(\frac{Z}{T}+\frac{1}{T_e}\right)^{-1}
  \Bigg\{
  \int_0^\infty\dd v
  \int_{B^{-1}_{0,{\rm max}}}^{B^{-1}}\dd\lambda
  \frac{v^3}{|v_{||}^{(0)}|}
  g_0^{\rm out}
  \nonumber\\[5pt]
 && +
  \int_I
  \dd v\,
   v^2
\int_{-\infty}^{\lambda_L(l)}\dd\lambda
\frac{
g_{{\rm rl}}}
{\sqrt{\lambda_r |\partial_l B_0(l_L)|(l-l_L)
-(\lambda - \lambda_r) B_0(l_L)
}}
 \nonumber\\[5pt]
&&  
  +
  \int_I
  \dd v\,
   v^2
\int_{-\infty}^{\lambda_R(l)}\dd\lambda
\frac{
g_{{\rm rl}}}
{\sqrt{\lambda_r |\partial_l B_0(l_R)|(l_R-l)
-(\lambda - \lambda_r) B_0(l_R)
}}
 \nonumber\\[5pt]
 && +
   \int_I
  \dd v\,
   v^2
\Bigg[
\frac{
1
}
{\sqrt{1-\lambda_r B_0(l)
}
} - \frac{1}{\sqrt{\lambda_r |\partial_l B_0(l_L)|(l -l_L)
}}
\nonumber\\[5pt]
&&
 - \frac{1}{\sqrt{\lambda_r |\partial_l B_0(l_R)|(l_R -l)
}}
\Bigg]
\int_{-\infty}^\infty\dd\lambda \,
g_{{\rm rl}}
\Bigg\}
,
\end{eqnarray}
where $l_L$ and $l_R$ denote the bounce points of the trajectory $\lambda = \lambda_r$,
\begin{equation}
\lambda_L(l) - \lambda_r = \frac{\lambda_r |\partial_l B_0(l_L)|(l-l_L)}{B_0(l_L)}
\end{equation}
and
\begin{equation}
\lambda_R(l) - \lambda_r = \frac{\lambda_r |\partial_l B_0(l_R)|(l_R-l)}{B_0(l_R)}.
\end{equation}
Expression \eq{eq:QN_sb-p_explicit} is useful in order to explain why we did not evaluate $J_\varphi^{(1)}$ at $\lambda_r$ in equation \eq{eq:DKEresonantlayer}. First, note that the first term in brackets in the fourth line of \eq{eq:QN_sb-p_explicit} diverges at $l = l_L$ and $l = l_R$. We distinguish two cases:

\begin{enumerate}
\item[(i)] If
 \begin{equation}
\left|\frac{2Z e \varphi'_0}{{m v^2  \lambda_r\partial_\lambda \overline{\partial_\rcoor B_0}(\lambda_{r}) }}\right| \gg \Delta_{{\rm sb-p}},
\end{equation}
 the dependence of $\lambda_r$ on $v$ is strong enough for the integral over $v$ to smooth out the divergence of $1/\sqrt{1-\lambda_r B_0(l)}$ at $l = l_L$ and $l = l_R$. Then,
 \begin{equation}
\varphi^{(1)} \sim \frac{T}{e}.
\end{equation}
 In this case, $\widehat{J_\varphi^{(1)}}$ can be replaced by $J_{\varphi, r}^{(1)}$ in \eq{eq:DKEresonantlayer}, where
\begin{equation}\label{eq:J1_varphi_r}
J_{\varphi,r}^{(1)}
  =
  -\frac{2Ze}{mv}
  \int_{l_{b_{10}}}^{l_{b_{20}}}
\frac{
\varphi^{(1)}
}
{\sqrt{1-\lambda_r B_0}}\, \dd l,
\end{equation}
and the solution of \eq{eq:DKEresonantlayer} can be found analytically, with $g_{\rm rl}$ approaching a Dirac delta distribution when the collision frequency tends to zero. This gives the standard superbanana-plateau regime in which the radial neoclassical fluxes are strictly independent of the collisionality. This regime was treated in \cite{Shaing2015} for tokamaks of finite aspect ratio with broken symmetry.
\item[(ii)] If
\begin{equation}\label{eq:sb-p_small_Er}
\left|\frac{2Z e \varphi'_0}{{m v^2  \lambda_r\partial_\lambda \overline{\partial_\rcoor B_0}(\lambda_{r}) }}\right|
 \ll \Delta_{{\rm sb-p}},
\end{equation}
then $\lambda_r$ is approximately independent of $v$ and no such smoothing happens. The effect of the accumulation of the resonances at the same $\lambda_r$ for every $v$ causes that, close to the bounce points of the resonant trajectory, the electrostatic potential become very large. Specifically, one finds that
\begin{equation}\label{eq:scaling1over6}
\varphi^{(1)} \sim
\begin{cases}
(\nu_*/\rho_*)^{-1/6} T/e & |l - l_j| \sim B_0 \Delta_{{\rm sb-p}} R_0\\
T/e  & |l - l_j| \gg B_0 \Delta_{{\rm sb-p}} R_0,
\end{cases}
\end{equation}
where $j = L, R$ (see figure \ref{fig:resonant_trajectory}). This subregime of the superbanana-plateau regime was discovered in \citet{CalvoPPCF2017} and we will focus on it in the next subsection, devoted to illustrating it numerically. When \eq{eq:sb-p_small_Er} holds, the radial neoclassical fluxes depend logarithmically on collisionality. The estimation \eq{eq:size_sb-p_layer} for the size of the layer also incorporates logarithmic corrections.
\end{enumerate}

Finally, we point out that, as clearly inferred from inspection of \eq{eq:DKEresonantlayer}, $g_{\rm rl}$ has no definite parity under stellarator symmetry transformations, and neither does $\varphi^{(1)}$.

\subsubsection{Superbanana-plateau regime in a large aspect ratio stellarator}
\label{sec:numerical_check_scaling1over6}

We start by giving some estimates that are needed below. First, note that $\partial_\lambda \partial_\rcoor J_r^{(0)} \sim \epsilon^{-3/2} v_t B_{00}$. Second, observe that for large aspect ratio $\partial_v\lambda_r \sim \epsilon / (B_{00} v_t)$ so that the term proportional to $\nu_v$ in \eq{eq:def_k} can be neglected compared to the term proportional to $\nu_\lambda$, giving $k\sim \epsilon^{3/2} \nu_\lambda B_{00}^{-2} R_0 \rho_*^{-1}$. Then, balancing the two terms on the left-hand side of the drift-kinetic equation \eq{eq:DKEresonantlayer}, we find that, for $\epsilon\ll 1$,
 \begin{equation}\label{eq:size_sb-p_layer_large_aspect_ratio}
 B_{00} \Delta_{{\rm sb-p}} \sim \epsilon\, \left(\frac{\nu_*}{\rho_*}\right)^{1/3}.
 \end{equation}
 
 For convenience, define $g_{{\rm rl}}[J_B^{(1)}]$ and $g_{{\rm rl}}[J_\varphi^{(1)}]$ as the solutions of
 \begin{eqnarray}
\partial_\lambda \partial_\rcoor J_r^{(0)}
\xi
\partial_\alpha  g_{\rm rl}[J_B^{(1)}]
 +
k \partial_\xi^2 g_{\rm rl}[J_B^{(1)}]
= 
\partial_\alpha J_{B,r}^{(1)}
\Upsilon F_M
\end{eqnarray}
and
 \begin{eqnarray}
\partial_\lambda \partial_\rcoor J_r^{(0)}
\xi
\partial_\alpha  g_{\rm rl}[J_\varphi^{(1)}]
 +
k \partial_\xi^2 g_{\rm rl}[J_\varphi^{(1)}]
= 
\partial_\alpha \widehat{J_\varphi^{(1)}}
\Upsilon F_M,
\end{eqnarray}
respectively. Of course, $g_{{\rm rl}} = g_{{\rm rl}}[J_B^{(1)}] + g_{{\rm rl}}[J_\varphi^{(1)}]$. Then,
 \begin{equation}
 g_{{\rm rl}}[J_B^{(1)}] \sim \left(\frac{\nu_*}{\rho_*}\right)^{-1/3}\frac{n}{v_t^3}
 \end{equation}
 and
  \begin{equation}
 g_{{\rm rl}}[J_\varphi^{(1)}] \sim \epsilon^{-1/2} \left(\frac{\nu_*}{\rho_*}\right)^{-1/3}\frac{J_\varphi^{(1)}}{v_t R_0}
 \frac{n}{v_t^3} \, .
 \end{equation}
 
\begin{figure}
\centering
\includegraphics[width=0.7\textwidth]{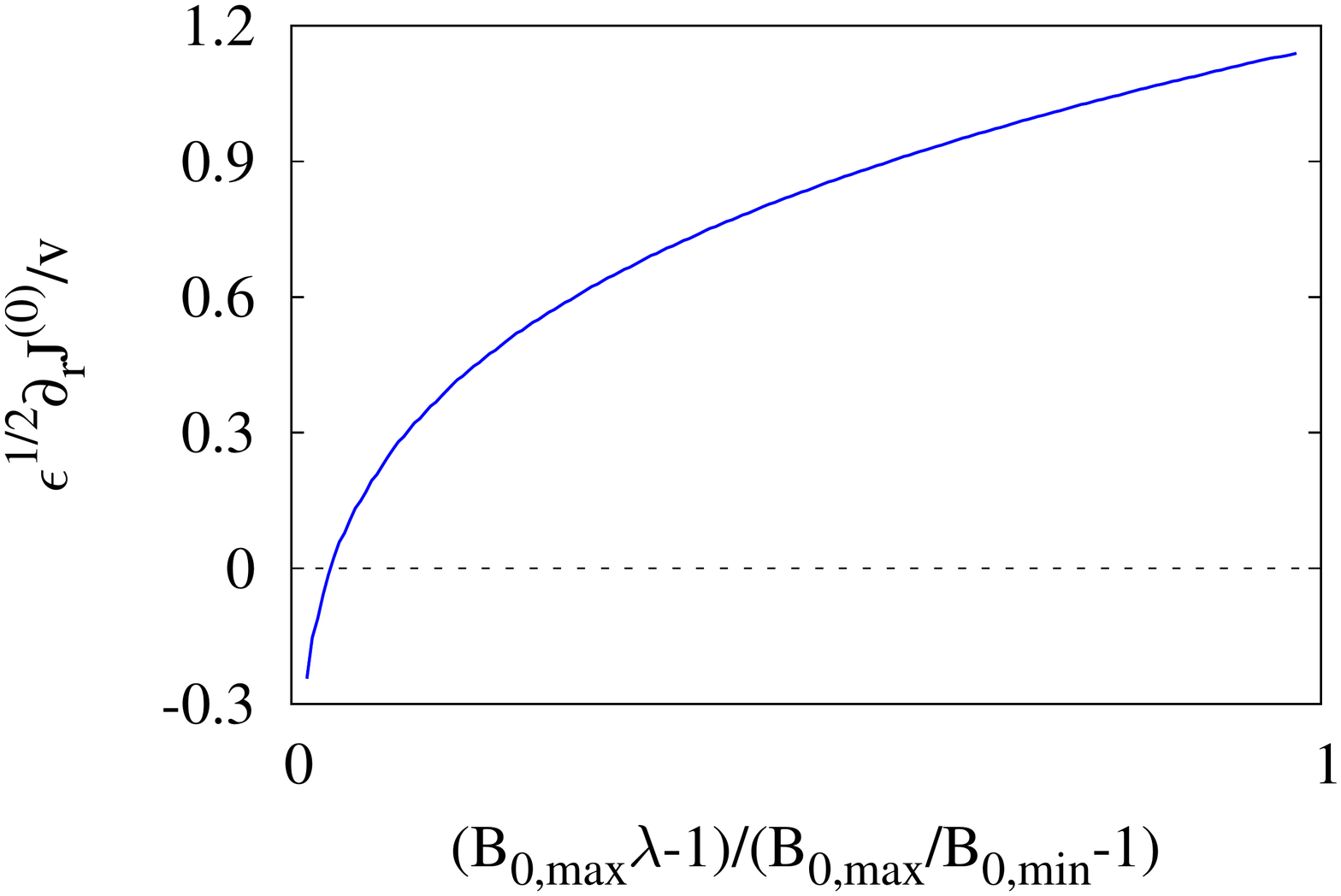}
\caption{$\partial_r J^{(0)}$ as a function of $\lambda$ in our model omnigeneous magnetic field, with $\epsilon = 1.1\times 10^{-4}$ and $\varphi'_0 = 0$. The value of $\lambda$ at which $\partial_r J^{(0)}$ vanishes determines the resonant value $\lambda_r$. Here, $(B_{0,{\rm max}} \lambda_r-1)/(B_{0, {\rm max}}/B_{0,{\rm min}}-1)=0.04$.}
\label{fig:drJ0}
\end{figure}

\begin{figure}
\centering
\includegraphics[width=0.325\textwidth]{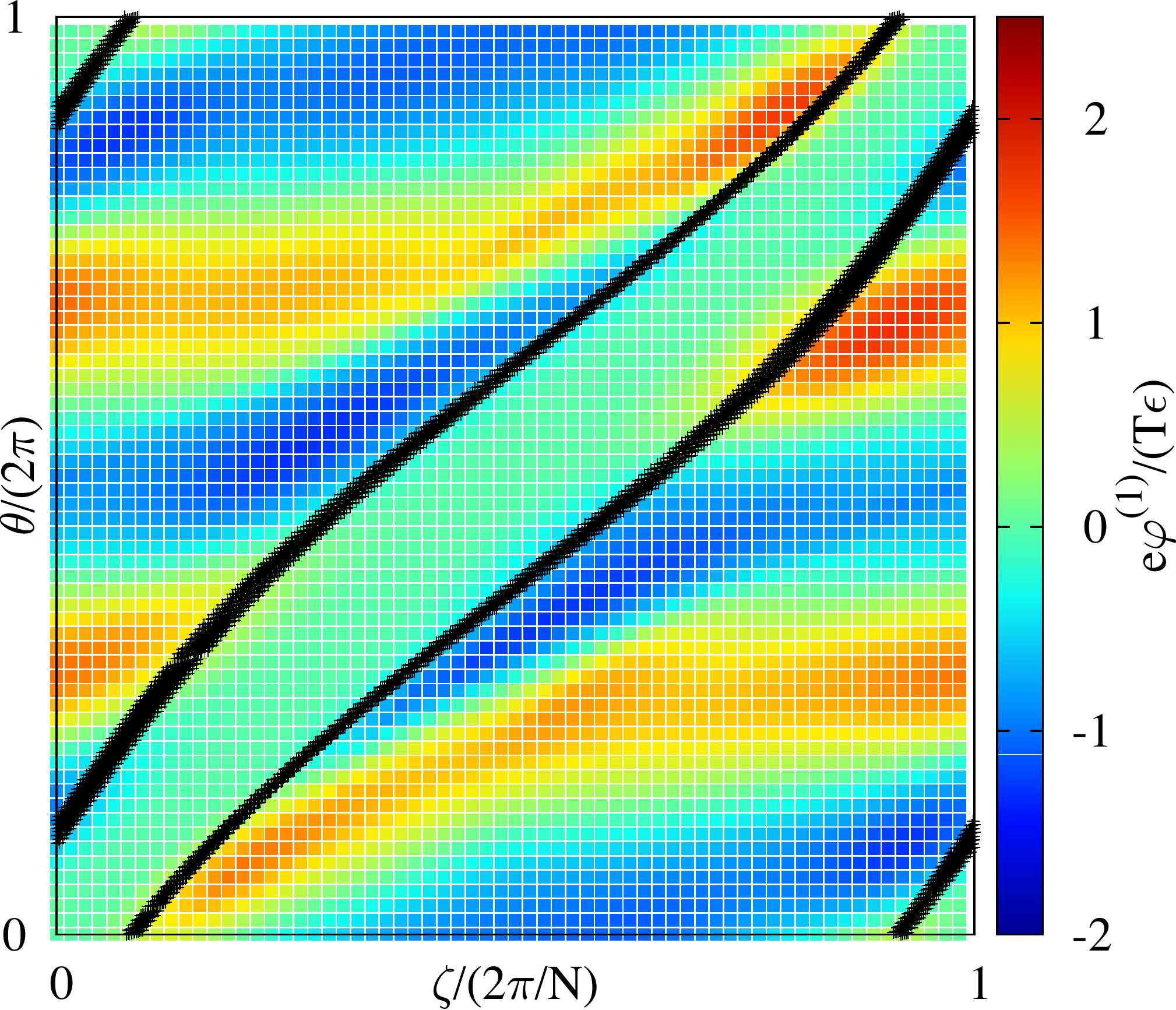}
\includegraphics[width=0.325\textwidth]{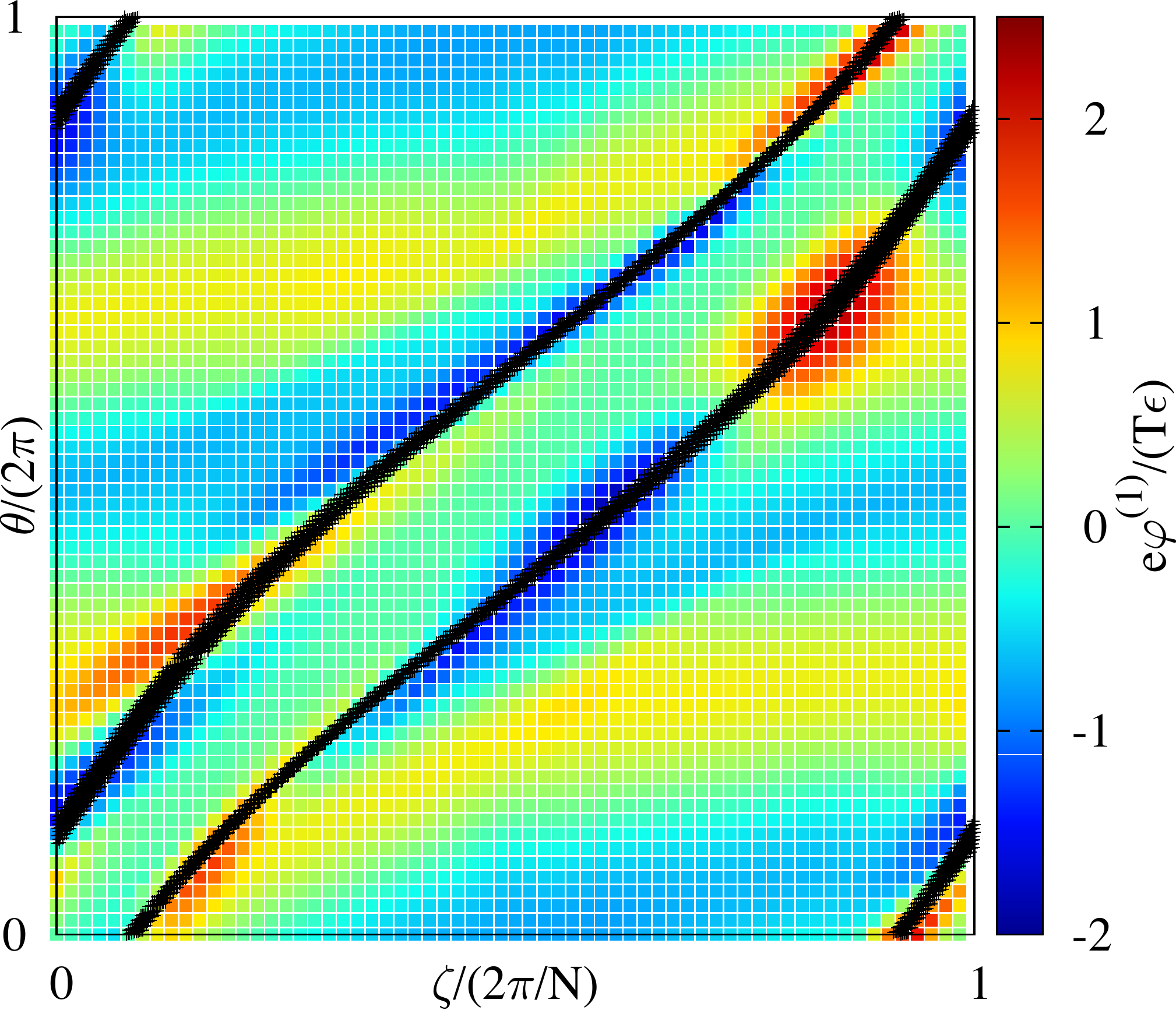}
\includegraphics[width=0.325\textwidth]{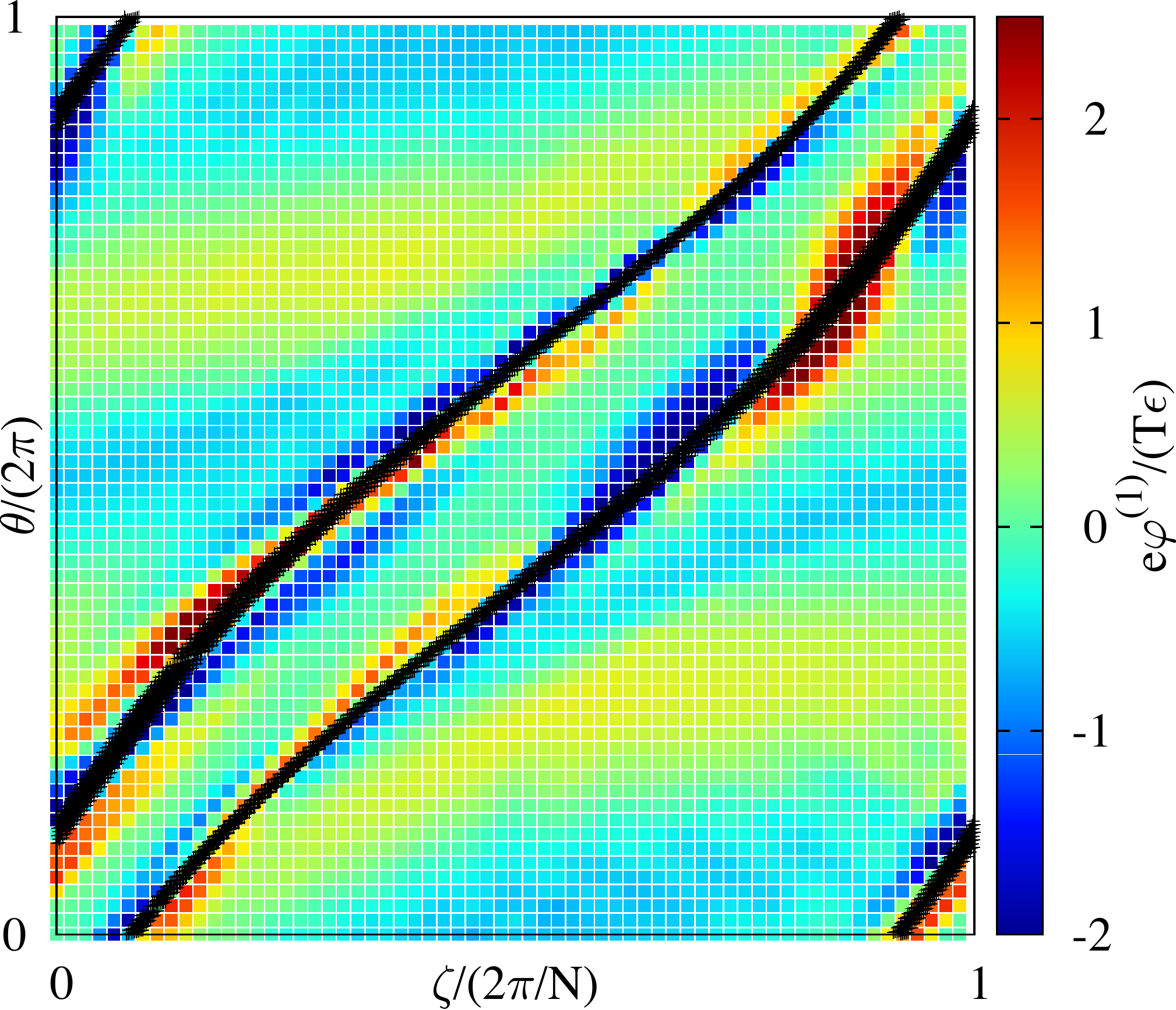}
\caption{Contour plot of $\varphi^{(1)}$ for $\epsilon = 1.1\times 10^{-4}$, $\varphi'_0 = 0$ and three values of the collisionality. From left to right, $\nu_*=1.2\times 10^{-12}$, $\nu_*=1.3\times 10^{-13}$, $\nu_*=1.3\times 10^{-14}$. The black lines are the two curves determined by the left and right bounce points of the resonant trajectory, $l_L$ and $l_R$, respectively. When $\nu_*$ decreases, the increase of $\varphi^{(1)}$ around $l_L$ and $l_R$, as predicted by \eq{eq:size_varphi_sb-p}, is confirmed. The color scale has been kept constant in the three figures, so that the increase in $\varphi^{(1)}$ becomes clearer; that is, in the center and right figures, $e|\varphi^{(1)}|/(T\epsilon)$ reaches larger values than the largest one indicated in the color scale bar.}
\label{fig:contour_plot_sb-p_regime}
\end{figure}

One can deal with the quasineutrality equation \eq{eq:QN_sb-p_explicit} by employing again the procedure followed after \eq{eq:def_intermediate_regime} and also in subsection \ref{sec:sqrtnu_small_Er}. The result is that
\begin{equation}\label{eq:size_varphi_sb-p}
\varphi^{(1)} \sim
\begin{cases}
(\nu_*/\rho_*)^{-1/6} \epsilon \, T/e & |l - l_j| \sim B_{00} \Delta_{{\rm sb-p}} R_0\\
\epsilon \, T/e & |l - l_j| \gg B_{00} \Delta_{{\rm sb-p}} R_0.
\end{cases}
\end{equation}

\begin{figure}
\centering
\includegraphics[width=0.8\textwidth]{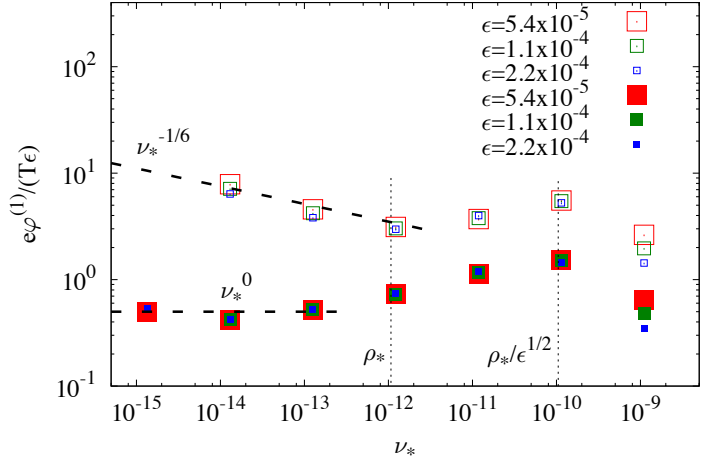}
\caption{Size of $\varphi^{(1)}$ as a function of $\nu_*$ and $\epsilon$. Here, $\varphi'_0 = 0$. The two scalings given in equation \eq{eq:size_varphi_sb-p} are checked. Full squares give the size of $\varphi^{(1)}$ at generic points, $|l - l_j| \gg B_{00} \Delta_{{\rm sb-p}} R_0$, whereas empty squares give the size of $\varphi^{(1)}$ at a point that is very close to a bounce point of the resonant trajectory, $|l - l_j| \sim B_{00} \Delta_{{\rm sb-p}} R_0$ (that is, empty squares give approximately the maximum of $\varphi^{(1)}$ on the flux surface).}
\label{fig:scaling_sb-p_regime}
\end{figure}

Let us check numerically the scalings of this subsection. First, in figure \ref{fig:drJ0}, we show the dependence of $\partial_r J^{(0)}$ on $\lambda$ for $\varphi'_0 = 0$ in our model magnetic configuration and we give, in particular, the location in phase space of the resonant value $\lambda_r$. In figure \ref{fig:contour_plot_sb-p_regime}, contour plots of $\varphi^{(1)}$ are shown for three different values of the collisionality, so that one can see how $\varphi^{(1)}$ becomes larger and larger around $l_L$ and $l_R$ as the collisionality decreases. Figure \ref{fig:scaling_sb-p_regime} provides a precise numerical check of the two scalings given in \eq{eq:size_varphi_sb-p}.

\section{Conclusions}
\label{sec:conclusions}

The component of the neoclassical electrostatic potential that is non-constant on flux surfaces, ${\tilde\varphi}$ (equivalently, the component of the electric field that is tangential to the surface), has been proven to have an important impact on stellarator radial impurity transport~\citep{GarciaPPCF2013}, triggering  interest in the correct calculation of the tangential electric field, which has often been neglected in stellarator neoclassical theory. On the numerical side, remarkable effort has been made during recent years (see \citet{GarciaNF2017} and references therein). In this article, we have studied analytically the scaling of ${\tilde\varphi}$ with collisionality and aspect ratio, and its structure on the flux surface.

The tangential electric field can be remarkably large when the collisionality is low enough for the tangential drifts to count in the drift-kinetic equation and especially when the tangential magnetic drift is non-negligible. If the collisionality and the radial electric field are such that the tangential magnetic drift cannot be dropped, the neoclassical equations can be kept linear and radially local only if the stellarator is sufficiently optimized~\citep{CalvoPPCF2017}; i.e. if the magnetic configuration is sufficiently close to omnigeneity~\citep{Cary1997b, Parra2015}. This is why we have conducted our study in the framework of a stellarator that is close to being omnigeneous, employing the techniques developed in \citet{CalvoPPCF2017}, where a small parameter $\delta$ exists that gives the size of the deviation from perfect omnigeneity. Throughout the paper, and employing some results of \citet{CalvoPPCF2017}, we discuss the calculation of the tangential electric field for collisionality values that cover the $1/\nu$, $\sqrt{\nu}$ and superbanana-plateau regimes, and in each of them provide its size, scaling with collisionality, aspect ratio and $\delta$, and behavior under stellarator symmetry transformations. In the treatment presented in \citet{CalvoPPCF2017} no assumption was made about the aspect ratio of the stellarator. Here, we have investigated how the results are modified when, apart from the $\delta$ expansion, one takes a subsidiary expansion in large aspect ratio. We have found new regimes that did not appear in tight aspect ratio devices. In particular, we have deduced that the maximum size admitted for ${\tilde\varphi} := \delta\varphi^{(1)}$ by the neoclassical equations is given by $\varphi^{(1)}\sim \epsilon T/e$, except at some special points where it can be larger. Table \ref{tab:table} summarizes the analytical results, providing the scaling with collisionality and aspect ratio in each regime. The notation is explained in Sections \ref{sec:basicequations} and \ref{sec:calculation_varphi1}.

The analytical results for each asymptotic regime have been verified and illustrated by numerical calculations of the code \texttt{KNOSOS}, that solves the equations derived in \citet{CalvoPPCF2017} for stellarators close to omnigeneity, the same set of equations that are the subject of study of this paper. In figure \ref{fig:varphi1_vs_nustar_realistic_values}, choosing realistic values for a stellarator reactor plasma, we have represented the size of $\varphi^{(1)}$ as a function of the collisionality, showing the different regimes in a single figure.

\begin{table}
  \begin{center}
    \begin{tabularx}{\linewidth}{L | L | L | L}
      \textbf{Parameter range} & \textbf{Size of $e\varphi^{(1)}/T$} & \textbf{Behavior under stellarator symmetry} & \textbf{Radial transport regime} \\
      \hline
      \multicolumn{4}{c}{Large radial electric field, $|\varphi'_0| \sim \epsilon^{-1}T/(e R_0)$} \\
      \hline
      $\epsilon^{-1}\rho_* \ll \nu_* \ll \epsilon^{3/2}$  & $\epsilon^{1/2}\rho_*/\nu_*$ & Antisymmetric & $1/\nu$\\[12pt]
      $\nu_* \ll \epsilon^{-1}\rho_*$  & $\epsilon^{3/2}$ & Symmetric & $\sqrt{\nu}$\\
      \hline
      \multicolumn{4}{c}{Small radial electric field, $|\varphi'_0| \lesssim T/(e R_0)$} \\
      \hline
      $\epsilon^{-1/2}\rho_* \ll \nu_* \ll \epsilon^{3/2}$  & $\epsilon^{1/2}\rho_*/\nu_*$ & Antisymmetric & $1/\nu$\\[12pt]
      $\rho_* \ll \nu_* \ll \epsilon^{-1/2}\rho_*$  & $\epsilon$ & Symmetric & $1/\nu$\\[12pt]
       $\nu_* \ll \rho_*$\newline ($\partial_r J^{(0)}$ has no zeros)  & $ \epsilon$ & Symmetric & $\sqrt{\nu}$
       \\[22pt]
       $\nu_* \ll \rho_*$\newline ($\partial_r J^{(0)}$ has zeros)  & $ \epsilon$, at generic points & No defined symmetry & Superbanana-plateau\\[-3pt]
       & $\epsilon (\nu_*/\rho_*)^{-1/6}$, at special points, if any  & & \\
    \end{tabularx}
  \end{center}
  \caption{Asymptotic regimes for $\varphi^{(1)}$}
    \label{tab:table}
  \end{table}

Finally, we comment on the expected effect of ${\tilde\varphi}$ on the radial neoclassical transport of the main ions. This effect is negligible compared to the transport due to $B_1$ except in regimes where ${\tilde\varphi}$ reaches its maximum size, given by $\varphi^{(1)}\sim \epsilon T/e$ (see Table \ref{tab:table}). In principle, in such regimes, the contributions of $B_1$ and ${\tilde\varphi}$ to the main ion neoclassical fluxes can be comparable. In practice, however, the maximum size reached by ${\tilde\varphi}$ is somewhat smaller. Obtaining $\varphi^{(1)}\sim \epsilon T/e$ seems to require density profile gradients $|n'/n| \sim (\epsilon R_0)^{-1}$, like the ones employed in this paper. If one takes a smaller density profile gradient and $|T'/T| \sim (\epsilon R_0)^{-1}$ (a situation that is more common in stellarators), the scaling of  ${\tilde\varphi}$ does not change but the actual value is smaller, due to the different way in which the density and temperature profiles enter the source term of the drift-kinetic equation.

\appendix
\section{Application of the Abel transform to integrals over trapped trajectories}
\label{sec:Abel_transform}

We give a useful inversion formula that interchanges the coordinates $l$ and $\lambda$. In this appendix, the coordinates $\rcoor$, $\alpha$ and $v$ are irrelevant, and we omit them.

For any function $f(l)$, we define $F(\lambda)$ as the following integral over the trapped trajectory determined by $\lambda\in[B_{0,{\rm max}}^{-1}, B_{0,{\rm min}}^{-1}]$,
\begin{equation}
F(\lambda) := \int_{l_{b_{10}}(\lambda)}^{l_{b_{20}}(\lambda)}\frac{f(l)}{\sqrt{1-\lambda B_0(l)}}\dd l.
\end{equation}
Then, given a value of the magnetic field ${\hat B}\in[B_{0,{\rm min}}, B_{0,{\rm max}}]$, we have
\begin{equation}\label{eq:inversion_formula}
\frac{f}{| \partial_l B_0 |}\Bigg\vert_{l_{b_{20}}({\hat B}^{-1})}
+
\frac{f}{| \partial_l B_0 |}\Bigg\vert_{l_{b_{10}}({\hat B}^{-1})}
=
-\frac{1}{\pi {\hat B}} \int_{{\hat B}^{-1}}^{B_{0, {\rm min}}^{-1}}
\frac{\dd F}{\dd \lambda}\frac{1}{\sqrt{\lambda {\hat B} - 1}} \dd\lambda,
\end{equation}
which can be proven by direct check. This formula can be viewed as a particularization of the Abel transform~\citep{Abel1826}.

For the sake of the application of the inversion formula in the main text of this article, we point out the following property, which is immediately derived from \eq{eq:inversion_formula}: if $f(l)$ is such that $ B_0(l_1) = B_0(l_2) \Rightarrow f(l_1) = f(l_2) $, then
\begin{equation}
F(\lambda) = 0, \mbox{ for every } \lambda \Rightarrow f(l) = 0, \mbox{ for every } l.
\end{equation}

\vspace{1cm}


The authors would like to thank an anonymous referee for pointing out that, in general, the energy diffusion piece of the collision operator cannot be neglected in equation \eq{eq:DKEresonantlayer}. This work has been carried out within the framework of the EUROfusion
Consortium and has received funding from the Euratom research and
training programme 2014-2018 under grant agreement No 633053. The
views and opinions expressed herein do not necessarily reflect those
of the European Commission. This research was supported in part by grant
 ENE2015-70142-P, Ministerio de Econom\'{\i}a y Competitividad,
Spain.

 \newcommand{\noop}[1]{}


\begin{thebibliography}{14}
\providecommand{\natexlab}[1]{#1}
\providecommand{\url}[1]{\texttt{#1}}
\expandafter\ifx\csname urlstyle\endcsname\relax
  \providecommand{\doi}[1]{doi: #1}\else
  \providecommand{\doi}{doi: \begingroup \urlstyle{rm}\Url}\fi
  
\bibitem[Abel (1826)]{Abel1826}
\textsc{Abel, N.~H.} 1826
\newblock Aufl\"osen einer mechanischen Aufgabe.
\newblock \emph{J. Reine Angew. Math.} {\bf 1}, 153.

\bibitem[Alonso \emph{et al.} (2016)]{Alonso2016}
\textsc{Alonso, J.~A., Velasco, J.~L., Calvo, I., Estrada, T., Fontdecaba, J.~M., García-Regaña, J.~M., Geiger, J., Landreman, M., McCarthy, K.~J., Medina, F., van
  Milligen, B.~Ph., Ochando, M.~A., Parra, F.~I., the TJ-II Team \& the W7-X Team} 2016
\newblock Parallel impurity dynamics in the TJ-II stellarator.
\newblock \emph{Plasma Phys. Control. Fusion} {\bf 58}, 074009.

\bibitem[Arévalo \emph{et al.} (2014)]{Arevalo2014}
\textsc{Arévalo, J., Alonso, J.~A., McCarthy, K.~J., Velasco, J.~L., García-Regaña, J.~M. \& Landreman, M.} 2014
\newblock Compressible impurity flow in the TJ-II stellarator.
\newblock \emph{Nucl. Fusion} {\bf 54}, 013008.

\bibitem[Beidler \emph{et al.} (2011)]{Beidler2011}
\textsc{Beidler, C.~D., Allmaier, K., Isaev, M.~Yu., Kasilov, S.~V., Kernbichler, W., Leitold, G.~O., Maassberg, H.,  Mikkelsen, D.~R., Murakami, S., Schmidt, M., Spong, D.~A., Tribaldos, V. and Wakasa, A.} 2011
\newblock Benchmarking of the mono-energetic transport coefficients---results from the International Collaboration on Neoclassical Transport in Stellarators (ICNTS).
\newblock \emph{Nucl. Fusion} {\bf 51}, 076001. 

\bibitem[Boozer (1981)]{Boozer81}
\textsc{Boozer, A.~H.} 1981
\newblock Plasma equilibrium with rational magnetic surfaces.
\newblock \emph{Phys. Fluids} {\bf 24}, 1999.

\bibitem[Burhenn \emph{et al.} (2009)]{BurhennNF2009}
\textsc{Burhenn, R., Feng Y., Ida, K., Maassberg, H., McCarthy, K.~J., Kalinina, D.,
  Kobayashi, M., Morita, S., Nakamura, Y., Nozato, H., Okamura, S., Sudo, S.,
  Suzuki, C., Tamura, N., Weller, A., Yoshinuma, M. \& Zurro B.} 2009
\newblock On impurity handling in high performance stellarator/heliotron
  plasmas.
\newblock \emph{Nucl. Fusion} {\bf 49}, 065005.

\bibitem[Calvo \emph{et al.} (2013)]{Calvo13}
\textsc{Calvo, I., Parra, F.~I., Velasco, J.~L. \& Alonso, J.~A.} 2013
\newblock Stellarators close to quasisymmetry.
\newblock {\em Plasma Phys. Control. Fusion} {\bf 55}, 125014.

\bibitem[Calvo \emph{et al.} (2014)]{Calvo14}
\textsc{Calvo, I., Parra, F.~I., Alonso, J.~A. \& Velasco, J.~L.} 2014
\newblock Optimizing stellarators for large flows.
\newblock {\em Plasma Phys. Control. Fusion} {\bf 56}, 094003.

\bibitem[Calvo \emph{et al.} (2015)]{Calvo15}
\textsc{Calvo, I., Parra, F.~I., Velasco, J.~L. \& Alonso, J.~A.} 2015
\newblock Flow damping in stellarators close to quasisymmetry.
\newblock {\em Plasma Phys. Control. Fusion} {\bf 57}, 014014.

\bibitem[Calvo \emph{et al.} (2017)]{CalvoPPCF2017}
\textsc{Calvo, I., Parra, F.~I., Velasco, J.~L. \& Alonso, J.~A.} 2017
\newblock The effect of tangential drifts on neoclassical transport in
  stellarators close to omnigeneity.
\newblock \emph{Plasma Phys. Control. Fusion} {\bf 59}, 055014.

\bibitem[Calvo \emph{et al.} (2018)]{Calvo2018}
\textsc{Calvo, I., Parra, F.~I., Velasco, J.~L., Alonso, J.~A. \& Garc\'{\i}a-Rega\~na, J.~M.}
\newblock Stellarator impurity flux driven by electric fields tangent to magnetic surfaces. \newblock \texttt{arXiv:1803.05691}.

\bibitem[Cary \& Shasharina (1997)]{Cary1997b}
\textsc{Cary, J.~R. \& Shasharina, S.~G.}
	1997
\newblock Omnigenity and quasihelicity in helical plasma confinement systems.
	\newblock{\em Phys. Plasmas}
	{\bf 4}, 3323.
	
\bibitem[Dewar \& Hudson (1998)]{Dewar1998}	
\textsc{Dewar, R.~L. \& Hudson, S.~R.} 1998
\newblock Stellarator symmetry.
\newblock \emph{Physica D: Nonlinear Phenomena} {\bf 112}, 275.
	
\bibitem[Galeev \& Sagdeev (1979)]{Galeev1979}
\textsc{Galeev, A.~A. \& Sagdeev, R.~Z.} 1979
\newblock Theory of neoclassical diffusion
\newblock{{\em Reviews of Plasma Physics}, vol 7, p 257}
\newblock ed Leontovich, M.~A.
\newblock (New York: Consultants Bureau).

\bibitem[García-Regaña \emph{et al.} (2013)]{GarciaPPCF2013}
\textsc{García-Regaña, J.~M., Kleiber, R., Beidler, C.~D., Turkin, Y., Maaßberg, H. \& Helander, P.} 2013
\newblock On neoclassical impurity transport in stellarator geometry.
\newblock \emph{Plasma Phys. Control. Fusion} {\bf 55}, 074008.

\bibitem[García-Regaña \emph{et al.} (2017)]{GarciaNF2017}
\textsc{García-Regaña, J.~M., Beidler, C.~D., Kleiber, R., Helander, P., Mollén, A., Alonso, J.~A., Landreman, M., Maaßberg, H., Smith, H.~M., Turkin, Y. \& Velasco, J.~L.} 2017
\newblock Electrostatic potential variation on the flux surface and its impact
  on impurity transport.
\newblock \emph{Nucl. Fusion} {\bf 57}, 056004.

\bibitem[García-Regaña \emph{et al.} (2018)]{Garcia2018}
\textsc{Garc\'ia-Rega\~na, J.~M., Estrada, T., Calvo, I., Velasco, J.~L., Alonso, J.~A., Carralero, D., Kleiber, R., Landreman, M., Moll\'en, A., S\'anchez, E., Slaby, C., TJ-II Team \& W7-X Team.}
\newblock On-surface potential and radial electric field variations in electron root stellarator plasmas.
\newblock \texttt{arXiv:1804.10424}.

\bibitem[Hall \& McNamara (1975)]{Hall1975}
\textsc{Hall, L.~S.  \& McNamara, B.} 1975
\newblock Three-dimensional equilibrium of the anisotropic, finite-pressure guiding-center plasma: Theory of the magnetic plasma.
\newblock{\em Phys. Fluids} {\bf 18}, 552.

\bibitem[Hazeltine (1973)]{Hazeltine73}
\textsc{Hazeltine, R.~D.} 1973
\newblock Recursive derivation of drift-kinetic equation.
\newblock{\em Plasma Phys.} {\bf 15}, 77.

\bibitem[Helander \& Sigmar (2002)]{Helander2002_book}
\textsc{Helander, P. \& Sigmar, D. J.} 2002 
\newblock Collisional Transport in Magnetized Plasmas (Cambridge
Monographs on Plasma Physics) ed Haines M G et al (Cambridge, UK: Cambridge University
Press)

\bibitem[Helander \& N\"uhrenberg (2009)]{Helander2009}
\textsc{Helander, P. \& N\"uhrenberg, J.} 2009
\newblock Bootstrap current and neoclassical transport in quasi-isodynamic stellarators.
\newblock \emph{Plasma Phys. Control. Fusion} {\bf 51}, 055004.

\bibitem[Helander \emph{et al.} (2017)]{HelanderPRL2017}
\textsc{Helander, P., Newton, S.~L., Moll\'en, A. \& Smith, H.~M.} 2017
\newblock Impurity transport in a mixed-collisionality stellarator plasma.
\newblock \emph{Phys. Rev. Lett.} {\bf 118}, 155002.

\bibitem[Ho \& Kulsrud (1987)]{Ho1987}
\textsc{Ho, D.~D.~M. \& Kulsrud, R.~M.} 1987
\newblock Neoclassical transport in stellarators.
\newblock{\em Phys. Fluids} {\bf 30}, 442.

\bibitem[Ida \emph{et al.} (2009)]{IdaPoP2009}
\textsc{Ida, K., Yoshinuma, M., Osakabe, M., Nagaoka, K., Yokoyama, M., Funaba, H.,
  Suzuki, C., Ido, T., Shimizu, A., Murakami, I., Tamura, N., Kasahara, H.,
  Takeiri, Y., Ikeda, K., Tsumori, K., Kaneko, O., Morita, S., Goto, M., Tanaka, K.,
  Narihara, K., Minami, T. \& Yamada, I.} 2009
\newblock Observation of an impurity hole in a plasma with an ion internal
  transport barrier in the large helical device.
\newblock \emph{Phys. Plasmas} {\bf 16}, 056111.

\bibitem[Klinger \emph{et al.} (2017)]{Klinger2017}
\textsc{Klinger, T., Alonso, A., Bozhenkov, S., Burhenn, R., Dinklage, A., Fuchert, G., Geiger, J., Grulke, O., Langenberg, A., Hirsch, M., Kocsis, G., Knauer, J., Kr\"amer-Flecken, A., Laqua, H., Lazerson, S., Landreman, M., Maassberg, H., Marsen, S., Otte, M., Pablant, N., Pasch, E., Rahbarnia, K., Stange, T., Szepesi, T., Thomsen, H., Traverso, P., Velasco, J.~L., Wauters, T., Weir, G., Windisch, T. and The Wendelstein 7-X Team} 2017
\newblock Performance and properties of the first plasmas of Wendelstein 7-X.
\newblock \emph{Plasma Phys. Control. Fusion} {\bf 59},  014018.

\bibitem[Kornilov \emph{et al.} (2005)]{Kornilov2005}
\textsc{Kornilov, V., Kleiber, R. \& Hatzky, R.} 2005
\newblock Gyrokinetic global electrostatic ion-temperature-gradient modes in finite $\beta$ equilibria of Wendelstein 7-X.
\newblock \emph{Nucl. Fusion} {\bf 45}, 238.

\bibitem[Landreman \& Catto (2012)]{Landreman2012}
\textsc{Landreman, M. \& Catto, P.~J.} 2012
\newblock Omnigenity as generalized quasisymmetry.
\newblock \emph{Phys. Plasmas} {\bf 19}, 056103.

\bibitem[Landreman \emph{et al.} (2014)]{Landreman2014}
\textsc{Landreman, M., Smith, H.~M., Moll\'en, A. \& Helander, P.} 2014
\newblock Comparison of particle trajectories and collision operators for collisional transport in nonaxisymmetric plasmas.
\newblock \emph{Phys. Plasmas} {\bf 21}, 042503. 

\bibitem[Matsuoka \emph{et al.} (2015)]{Matsuoka2015}
\textsc{Matsuoka, S., Satake, S., Kanno, R. \& Sugama, H.} 2015
\newblock Effects of magnetic drift tangential to magnetic surfaces on neoclassical transport in non-axisymmetric plasmas.
\newblock \emph{Phys. Plasmas} {\bf 22}, 072511.

\bibitem[McCormick \emph{et al.} (2002)McCormick, Grigull, Burhenn, Brakel, Ehmler,
  Feng, Gadelmeier, Giannone, Hildebrandt, Hirsch, Jaenicke, Kisslinger,
  Klinger, Klose, Knauer, K\"onig, K\"uhner, Laqua, Naujoks, Niedermeyer,
  Pasch, Ramasubramanian, Rust, Sardei, Wagner, Weller, Wenzel, and
  Werner]{McCormickPRL2002}
\textsc{McCormick, K., Grigull, P., Burhenn, R., Brakel, R., Ehmler, H., Feng, Y.,
  Gadelmeier, F., Giannone, L., Hildebrandt, D., Hirsch, M., Jaenicke, R.,
  Kisslinger, J., Klinger, T., Klose, S., Knauer, J.~P., K\"onig, R., K\"uhner, G., Laqua, H.~P., Naujoks, D., Niedermeyer, H., Pasch, E., Ramasubramanian, N., Rust, N., Sardei, F., Wagner, F., Weller, A., Wenzel, U. \& Werner, A.} 2002
\newblock New advanced operational regime on the W7-AS stellarator.
\newblock \emph{Phys. Rev. Lett.} {\bf 89}, 015001.

\bibitem[Mollén \emph{et al.} (2018)]{Mollen2018}
\textsc{Mollén, A., Landreman, M., Smith, H.~M., García-Regaña, J.~M., \& Nunami, M.} 2018
\newblock Flux-surface variations of the electrostatic potential in stellarators: impact on the radial electric field and neoclassical impurity transport.
\newblock {\em Plasma Phys. Control. Fusion} {\bf 60}, 084001.
\bibitem[Mynick (1984)]{Mynick1984}
\textsc{Mynick, H.~E.} 1984
\newblock Calculation of the poloidal ambipolar field in a stellarator and its effect on transport.
\newblock {\em Phys. Fluids} {\bf 27}, 2086.

\bibitem[Parra \& Calvo (2011)]{Parra11}
\textsc{Parra, F.~I. \& Calvo. I.} 2011
\newblock Phase-space Lagrangian derivation of electrostatic gyrokinetics in general geometry.
\newblock {\em Plasma Phys. Control. Fusion} {\bf 53}, 045001.

\bibitem[Parra \emph{et al.} (2015)]{Parra2015}
\textsc{Parra, F.~I., Calvo, I., Helander, P. and Landreman, M.} 2015
\newblock Less constrained omnigeneous stellarators.
\newblock{\em Nucl. Fusion} {\bf 55}, 033005.

\bibitem[Paul \emph{et al.} (2017)]{Paul2017}
\textsc{Paul, E.~J., Landreman, M., Poli, F.~M., Spong, D.~A., Smith, H.~M. \& Dorland, W.} 2017
\newblock Rotation and neoclassical ripple transport in ITER.
\newblock {\em Nucl. Fusion} {\bf 57}, 116044.

\bibitem[Pedrosa \emph{et al.} (2015)]{PedrosaNF2015}
\textsc{Pedrosa, M.~A., Alonso, J.~A., García-Regaña, J.~M., Hidalgo, C., Velasco, J.~L.,
  Calvo, I., Kleiber, R., Silva, C. \& Helander, P.} 2015
\newblock Electrostatic potential variations along flux surfaces in
  stellarators.
\newblock \emph{Nucl. Fusion} {\bf 55}, 052001.

\bibitem[Shaing (2015)]{Shaing2015}
\textsc{Shaing, K.~C.} 2015
\newblock Superbanana and superbanana plateau transport in finite aspect ratio tokamaks with broken symmetry.
\newblock {\em J. Plasma Physics} {\bf 81}, 905810203.

\bibitem[Velasco \emph{et al.} (2017)]{VelascoNF2017}
\textsc{Velasco, J.~L., Calvo, I., Satake, S., Alonso, J.~A., Nunami, M., Yokoyama, M., Sato, M.,
  Estrada, T., Fontdecaba, J.~M., Liniers, M., McCarthy, K.~J., Medina, F., van
  Milligen, B.~Ph., Ochando, M., Parra F.~I., Sugama, H., Zhezhera, A., the LHD experimental
  team \& the TJ-II team} 2017
\newblock Moderation of neoclassical impurity accumulation in high temperature
  plasmas of helical devices.
\newblock \emph{Nucl. Fusion} {\bf 57}, 016016.

\bibitem[Velasco \emph{et al.} (2018)]{Velasco2018}
\textsc{Velasco, J.~L., Calvo, I., Garc\'{\i}a-Rega\~na, J.~M., Parra, F.~I., Satake, S., Alonso, J.~A. \& the LHD team} 2018
\newblock Large tangential electric fields in plasmas close to temperature screening.
\newblock \emph{Plasma Phys. Control. Fusion} {\bf 60}, 074004.

\end{thebibliography}
\end{document}